\documentclass[submission,copyright,creativecommons]{eptcs}
\providecommand{\event}{DCM 2023} 

\usepackage{iftex}

\ifpdf
  \usepackage{underscore}         
  \usepackage[T1]{fontenc}        
\else
  \usepackage{breakurl}           
\fi

\usepackage{amsmath,amsthm,amssymb,float,mathtools}
\usepackage[american]{babel}
\usepackage{bussproofs}
\usepackage{color}
\usepackage{crossreftools}
\usepackage{enumitem}
\usepackage{multibib}
\usepackage{MnSymbol}
\usepackage{soul}
\usepackage{stmaryrd}
\usepackage{xspace}

\usepackage{tikz,fp,tikz-cd,tkz-graph}
\usepackage{pgfplots}
\usetikzlibrary{arrows.meta,arrows,calc,automata,angles,quotes,matrix,
                positioning,intersections,through,shapes.geometric,shapes.symbols,shapes.multipart,through,trees,
                decorations.markings,decorations.text,
                }                

\makeatletter
\def\calcLength(#1,#2)#3{%
\pgfpointdiff{\pgfpointanchor{#1}{center}}%
             {\pgfpointanchor{#2}{center}}%
\pgf@xa=\pgf@x%
\pgf@ya=\pgf@y%
\FPeval\@temp@a{\pgfmath@tonumber{\pgf@xa}}%
\FPeval\@temp@b{\pgfmath@tonumber{\pgf@ya}}%
\FPeval\@temp@sum{(\@temp@a*\@temp@a+\@temp@b*\@temp@b)}%
\FProot{\FPMathLen}{\@temp@sum}{2}%
\FPround\FPMathLen\FPMathLen5\relax
\global\expandafter\edef\csname #3\endcsname{\FPMathLen}
}
\makeatother   

\pgfdeclarelayer{background}
\pgfdeclarelayer{foreground}
\pgfsetlayers{background,main,foreground}

\tikzstyle{->>>} =
  [shorten >= 0.3mm,
   decoration =
     {markings,
      mark=at position -2.4mm with {\arrow{>}},
      mark=at position -1.2mm with {\arrow{>}},
      mark=at position 1 with {\arrow{>}}},
   postaction={decorate}]

\pgfdeclaredecoration{funbisim}{final}{
  \state{final}[width=\pgfdecoratedpathlength]{
    \draw[->]
      (0,\pgfdecorationsegmentamplitude+0.1mm) -- +(\pgfdecoratedpathlength,0);
    \draw[-]
      (0,-\pgfdecorationsegmentamplitude-0.1mm) -- +(\pgfdecoratedpathlength,0);}}
\tikzset{
  funbisim/.style={
    decoration={funbisim, amplitude=0.25ex},
    decorate,
    funbisim options/.style={#1}    
  }}

\pgfdeclaredecoration{bisim}{final}{
  \state{final}[width=\pgfdecoratedpathlength]{
    \draw[<->]
      (0,\pgfdecorationsegmentamplitude+0.1mm) -- +(\pgfdecoratedpathlength,0);
    \draw[-]
      (0,-\pgfdecorationsegmentamplitude-0.1mm) -- +(\pgfdecoratedpathlength,0);}}
\tikzset{
  bisim/.style={
    decoration={bisim, amplitude=0.25ex},
    decorate,
    bisim options/.style={#1}    
  }}

\makeatletter
\pgfdeclareshape{transbox}{
  \inheritsavedanchors[from=rectangle]
  \inheritanchor[from=rectangle]{center}
  \inheritanchor[from=rectangle]{north}
  \inheritanchor[from=rectangle]{south}
  \inheritanchor[from=rectangle]{west}
  \inheritanchor[from=rectangle]{east}
  \inheritbehindbackgroundpath[from=rectangle]
  \inheritbackgroundpath[from=rectangle]
  \inheritbeforebackgroundpath[from=rectangle]
  \inheritbehindforegroundpath[from=rectangle]
  \inheritforegroundpath[from=rectangle]
  \inheritbeforeforegroundpath[from=rectangle]
}
\makeatother

\title{From Compactifying Lambda-Letrec Terms\\
        to Recognizing Regular-Expression Processes
        \\
       {\small (Extended Abstract and Literature)}}

\author{Clemens Grabmayer
\institute{Department of Computer Science\\Gran Sasso Science Institute\\ L'Aquila, Italy}\\
\email{clemens.grabmayer@gssi.it}}
\def\titlerunning{From Compactifying Lambda-Letrec Terms
                  to Recognizing Regular-Expression Processes}
\def\authorrunning{C. Grabmayer}

\theoremstyle{plain}
\newtheorem{resque}{Research Question}
\newtheorem{respro}[resque]{Research Project}
\newtheorem*{respro*}{Research Project}

\theoremstyle{definition}
\definecolor{azure}{rgb}{0.94,1.00,1.00}
\definecolor{brown}{rgb}{.75,.25,.25}
\definecolor{cyan}{rgb}{0.25,0.88,0.82}
\definecolor{chocolate}{rgb}{0.82,0.41,0.12}
\definecolor{darkcyan}{rgb}{0.5,0,1}
\definecolor{darkgreen}{rgb}{0,0.39,0}
\definecolor{darkmagenta}{rgb}{0.5,0,0.5}
\definecolor{darkgoldenrod}{RGB}{184,134,11}
\definecolor{firebrick}{RGB}{175,25,25}
\definecolor{forestgreen}{rgb}{0.13,0.55,0.13}
\definecolor{goldenrod}{RGB}{218,165,32}
\definecolor{grey}{RGB}{72,72,72}
\definecolor{lightcyan}{rgb}{0.88,1.00,1.00}
\definecolor{lightpink}{rgb}{1.00,0.71,0.76}
\definecolor{myyellow}{RGB}{235,235,0}
\definecolor{lightyellow}{rgb}{1.00,1.00,0.88}
\definecolor{lightgoldenrod}{rgb}{0.83,0.97,0.51}
\definecolor{lightgoldenrodyellow}{rgb}{0.98,0.98,0.82}
\definecolor{lightskyblue}{rgb}{0.53,0.81,0.98}
\definecolor{moccasin}{rgb}{1.00,0.89,0.71}
\definecolor{magenta}{rgb}{1,0,1}
\definecolor{navyblue}{rgb}{0,0,0.5}
\definecolor{orange}{rgb}{1.0,0.65,0.0}
\definecolor{orangered}{rgb}{1.0,0.27,0.0}
\definecolor{palegreen}{rgb}{0.60,0.98,0.60}
\definecolor{powderblue}{rgb}{0.69,0.88,0.90}
\definecolor{purple}{rgb}{1,0.5,1}
\definecolor{royalblue}{RGB}{65,105,225}
\definecolor{mediumblue}{RGB}{0,0,205}
\definecolor{cornflowerblue}{RGB}{100,149,237}
\definecolor{springgreen}{rgb}{0.0,1.0,0.5}
\definecolor{turquoise}{rgb}{0.25,0.88,0.82}
\definecolor{snow}{rgb}{1.00,0.98,0.98}
\definecolor{tan}{rgb}{0.82,0.71,0.55}
\definecolor{red}{rgb}{1,0,0}
\definecolor{violetred}{RGB}{208,32,144}

\newcommand{\colorin}[1]{\textcolor{#1}}

\newcommand{\black}{\colorin{black}}

\newcommand{\chocolate}{\colorin{chocolate}}
\newcommand{\colorred}{\colorin{red}}

\newcommand{\darkcyan}{\colorin{darkcyan}}

\newcommand{\forestgreen}{\colorin{forestgreen}}
\newcommand{\firebrick}{\colorin{firebrick}}

\newcommand{\magenta}{\colorin{magenta}}
\newcommand{\mediumblue}{\colorin{mediumblue}}

\newcommand\inMath[1]{\ensuremath{#1}\xspace}

\newcommand{\nb}{\nobreakdash}

\newcommand{\nf}{\normalfont}
  \newcommand{\textnf}[1]{\text{\nf{#1}}}

\newcommand{\punc}[1]{\ensuremath{\hspace*{1.5pt}{#1}}}

\newcommand{\tightfbox}[1]{{\fboxsep=1.5pt\fbox{#1}}}

\newcommand{\funin}{\mathrel{:}}
\newcommand{\fap}[2]{{#1}(\hspace*{-0.5pt}{#2}\hspace*{-0.5pt})}
\newcommand{\funap}[2]{#1({#2})}

\newcommand{\iap}[2]{#1_{#2}}
\newcommand{\bap}[2]{#1_{#2}}
\newcommand{\pap}[2]{#1^{#2}}
\newcommand{\bpap}[3]{#1_{#2}^{#3}}

\newcommand{\sidfunon}{\iap{\textrm{\nf id}}}

\newcommand{\sdefdby}{{:=}}
\newcommand{\defdby}{\mathrel{\sdefdby}}

\newcommand{\overlinebar}[1]{\mathbf{\overline{\text{$#1$}}}}

\newcommand{\niks}{}

\newcommand{\eqcl}[1]{\iap{\left[{#1}\right]}}

\newcommand{\FPT}{\textnf{FPT}}  
  
\newcommand{\setexp}[1]{\left\{{#1}\right\}}

\newcommand{\stexpzero}{0}
\newcommand{\stexpone}{1}
\newcommand{\aact}{a}
\newcommand{\bact}{b}
\newcommand{\cact}{c}

\newcommand{\astexp}{e}
\newcommand{\bstexp}{f}
\newcommand{\cstexp}{g}

\newcommand{\astexpi}{\iap{\astexp}}

\newcommand{\astexpacc}{\astexp'}

\newcommand{\astexpacci}{\iap{\astexpacc}}

\newcommand{\sstexpsum}{+}
\newcommand{\stexpsum}[2]{{#1\sstexpsum #2}}
\newcommand{\sstexpprod}{\cdot}
\newcommand{\stexpprod}[2]{{#1\sstexpprod #2}}
\newcommand{\sstar}{*}

\newcommand{\stexpit}[1]{{#1 ^{\sstar}}}

\newcommand{\sstexpbit}{\circledast} 
\newcommand{\stexpbit}[2]{{#1}\hspace*{0.35pt}\pap{}{\sstexpbit}\hspace*{-0.6pt}{#2}}

\newcommand{\milnersys}{\text{\nf\sf Mil}}  

\newcommand{\coindmilnersys}{\text{\nf\sf cMil}}

\newcommand{\sprocint}{P} 
\newcommand{\procint}{\fap{\sprocint}}
\newcommand{\sprocintone}{\underline{\sprocint}} 
\newcommand{\procintone}{\fap{\sprocintone}}

\newcommand{\sprocsem}{P} 
\newcommand{\procsem}[1]{\llbracket{#1}\rrbracket_{\hspace*{-0.5pt}\sprocsem}}

\newcommand{\sprocsemeq}{\iap{=}{{\scriptscriptstyle \procsem{\cdot}}}}
\newcommand{\procsemeq}{\mathrel{\sprocsemeq}}
\newcommand{\snotprocsemeq}{\iap{\neq}{{\scriptscriptstyle \procsem{\cdot}}}}
\newcommand{\notprocsemeq}{\mathrel{\snotprocsemeq}}

\newcommand{\langsemexpressible}{$\langsem{\cdot}$\nb-expressible}

\newcommand{\procintexpressible}{$\sprocint$\nb-expressible}
\newcommand{\procintexpressibility}{$\sprocint$\nb-ex\-pres\-si\-bi\-lity}
\newcommand{\procsemexpressible}{$\procsem{\cdot}$\nb-expressible}
\newcommand{\procsemexpressibility}{$\procsem{\cdot}$\nb-ex\-pres\-si\-bi\-lity}

\newcommand{\slangint}{L} 
\newcommand{\langint}{\fap{\slangint}} 
\newcommand{\slangsem}{L} 
\newcommand{\langsem}[1]{\llbracket{#1}\rrbracket_{\slangsem}}

\newcommand{\slangsemeq}{{\iap{=}{\scriptscriptstyle \langsem{\cdot}}}} 
\newcommand{\langsemeq}{\mathrel{\slangsemeq}}

\newcommand{\sformeq}{{=}}
\newcommand{\formeq}{\mathrel{\sformeq}}

\newcommand{\agraph}{G}
\newcommand{\agraphacc}{\agraph'}
\newcommand{\agraphi}{\iap{\agraph}}
\newcommand{\agraphbp}{\bpap{\agraph}}

\newcommand{\aloop}{\textit{LG}}
\newcommand{\aloopi}{\iap{\aloop}}

\newcommand{\selimred}{{\Rightarrow_{\textnf{elim}}}}
\newcommand{\elimred}{\mathrel{\selimred}}
\newcommand{\selimredrtc}{{\bpap{\Rightarrow}{\textnf{elim}}{*}}}
\newcommand{\elimredrtc}{\mathrel{\selimredrtc}}

\newcommand{\picarrowstart}{\raisebox{2pt}{\begin{tikzpicture}%
                                             \draw[<-,very thick,>=latex,chocolate,shorten <=2pt](0,0) -- ++ (180:{12pt});%
                                           \end{tikzpicture}}}
  
\newcommand{\pictermvert}{\begin{tikzpicture}%
                           \node[draw,chocolate,very thick,circle,minimum width=2.5pt,fill,inner sep=0pt,outer sep=2pt](v){};%
                           \draw[thick,chocolate] (v) circle (0.12cm);%
                         \end{tikzpicture}}  
                                        
\newcommand{\picbisimlink}{\begin{tikzpicture}
                             \node(v){};
                             \draw[magenta,densely dashed] (v) to ($(v) + (0.6cm,0cm)$);
                           \end{tikzpicture}}

\newcommand{\sLEE}{\text{\nf LEE}}
\newcommand{\sLLEE}{\text{\nf LLEE}}

\newcommand{\LEE}{\sLEE}

\newcommand{\LLEE}{\sLLEE}

\newcommand{\start}{\averti{\hspace*{-0.5pt}\text{\nf s}}}

\newcommand{\actions}{A}

\newcommand{\slt}[1]{{\xrightarrow{#1}}}

\newcommand{\lt}[1]{\mathrel{\slt{#1}}}

\newcommand{\sterminates}{\Downarrow}
\newcommand{\terminates}[1]{{#1}{\sterminates}}
\newcommand{\snotterminates}{\nDownarrow}
\newcommand{\notterminates}[1]{{#1}{\snotterminates}}

\newcommand{\asettrans}{T}

\newcommand{\saTSS}{{\cal{T}}}
\newcommand{\StExpTSS}{\text{$\saTSS$}}

\newcommand{\avert}{v}
\newcommand{\bvert}{w}

\newcommand{\averti}{\iap{\avert}}
\newcommand{\bverti}{\iap{\bvert}}

\newcommand{\bvertbar}{\overlinebar{\bvert}}

\newcommand{\bvertbari}{\iap{\bvertbar}}

\newcommand{\sone}{\firebrick{1}}
 
\newcommand{\scpfun}{{\textit{cp}}}

\newcommand{\sbisimsubscript}{
    \setbox0=\hbox{\kern-.1ex{$\leftrightarrow$}\kern-.1ex}
    \setbox1=\vbox{\hbox{\raise .1ex \box0}\hrule}%
    \inMath{\mathrel{\hbox{\scalebox{0.75}{\box1}}}}
  }
  
\newcommand{\sfunonebisim}{%
    \setbox0=\hbox{\kern-.1ex{\magenta{$\rightarrow$}}\kern-.1ex}
    \setbox1=\vbox{\hbox{\raise .1ex \box0}\hrule}%
    \ensuremath{\magenta{\hbox{\kern.05ex\box1\kern.1ex}}}
  }

\newcommand{\sconvfunonebisim}{%
    \setbox0=\hbox{\kern-.1ex{$\leftarrow$}\kern-.1ex}
    \setbox1=\vbox{\hbox{\raise .1ex \box0}\hrule}%
    \ensuremath{\magenta{\hbox{\kern.05ex\box1\kern.1ex}}}
  }

\newcommand{\sonebisim}{%
    \setbox0=\hbox{\kern-.1ex{$\leftrightarrow$}\kern-.1ex}
    \setbox1=\vbox{\hbox{\raise .1ex \box0}\hrule}%
    \ensuremath{\magenta{\hbox{\kern.1ex\box1\kern.1ex}}}
  }
\newcommand{\onebisim}{\mathrel{\sonebisim}}

\newcommand{\sfunbisim}{%
    \setbox0=\hbox{\kern-.1ex{$\rightarrow$}\kern-.1ex}
    \setbox1=\vbox{\hbox{\raise .1ex \box0}\hrule}%
    {\hbox{\kern.05ex\box1\kern.1ex}}
  }
\newcommand{\funbisim}{\hspace*{-0pt}\mathrel{\sfunbisim}}

\newcommand{\sshaftfunbisim}{
    \setbox0=\hbox{\kern-.1ex{{---}}\kern-.1ex}
    \setbox1=\vbox{\hbox{\raise .1ex \box0}\hrule}%
    {\hbox{\kern.05ex\box1\kern.1ex}}
  }

\newcommand{\sconvfunbisim}{%
    \setbox0=\hbox{\kern-.1ex{$\leftarrow$}\kern-.1ex}
    \setbox1=\vbox{\hbox{\raise .1ex \box0}\hrule}%
    {\hbox{\kern.05ex\box1\kern.1ex}}
  }

\newcommand{\sbisim}{%
    \setbox0=\hbox{\kern-.1ex{$\leftrightarrow$}\kern-.1ex}
    \setbox1=\vbox{\hbox{\raise .1ex \box0}\hrule}%
    \hbox{\kern.1ex\box1\kern.1ex}
  }
\newcommand{\bisim}{\mathrel{\sbisim\hspace*{1pt}}}

\newcommand{\sfunonebisimvia}[1]{%
    \setbox0=\hbox{\kern-.1ex{$\rightarrow$}\kern-.1ex}
    \setbox1=\vbox{\hbox{\raise .1ex \box0}\hrule}%
    {\bap{\magenta{\hbox{\kern.05ex\box1\kern.1ex}}}{#1}}
  }

\newcommand{\sconvfunonebisimvia}[1]{%
    \setbox0=\hbox{\kern-.1ex{$\leftarrow$}\kern-.1ex}
    \setbox1=\vbox{\hbox{\raise .1ex \box0}\hrule}%
    {\bap{\magenta{\hbox{\kern.05ex\box1\kern.1ex}}}{#1}}
  }

\newcommand{\sonebisimvia}[1]{%
    \setbox0=\hbox{\kern-.1ex{$\leftrightarrow$}\kern-.1ex}
    \setbox1=\vbox{\hbox{\raise .1ex \box0}\hrule}%
    {\bap{\magenta{\hbox{\kern.05ex\box1\kern.1ex}}}{\hspace*{-1.5pt}#1}}
  }

\newcommand{\assocstexpsum}{\textnf{(A1)}}
\newcommand{\neutralstexpsum}{\textnf{(A2)}}
\newcommand{\commstexpsum}{\textnf{(A3)}}
\newcommand{\idempotstexpsum}{\textnf{(A4)}}
\newcommand{\assocstexpprod}{\textnf{(A5)}}
\newcommand{\rightdistr}{\textnf{(A6)}}
\newcommand{\leftidstexpprod}{\textnf{(A7)}}
\newcommand{\rightidstexpprod}{\textnf{(A8)}}
\newcommand{\deadlockax}{\textnf{(A9)}}
\newcommand{\recdefstexpit}{\textnf{(A10)}}
\newcommand{\termbodystexpit}{\textnf{(A11)}}

\newcommand{\sRSP}{\textrm{\nf RSP}}
\newcommand{\RSPstar}{\text{$\sRSP^{*}\hspace*{-1pt}$}}
\newcommand{\USP}{\textrm{\nf USP}}

\newcommand{\blambda}{\boldsymbol{\lambda}} 

\newcommand{\lambdaletreccal}{\inMath{\iap{\blambda}{\txtletrec}}}

\newcommand{\inflambdacal}{\inMath{\boldsymbol{\lambda}^{\hspace*{-1pt}\boldsymbol{\infty}}}}

\newcommand{\lambdamucal}{\inMath{\boldsymbol{\lambda}_{\hspace*{-1.5pt}\boldsymbol{\mu}}}} 

\newcommand{\letin}[2]{\txtlet\;{#1}\;\txtin\;{#2}}

\newcommand{\txtin}{{\text{\sf in}}}
\newcommand{\txtlet}{{\text{\sf let}}}
\newcommand{\txtletrec}{\text{\normalfont\sf letrec}}

\newcommand{\txtmu}{\ensuremath{\mu}}

\newcommand{\txtinflambdacal}{\ensuremath{\lambda^{\hspace*{-1pt}\infty}}}

\newcommand{\txtlambdaletreccal}{\ensuremath{\iap{\lambda}{\txtletrec}}}

\newcommand{\sslabs}{\lambda}
\newcommand{\slabs}[1]{\sslabs{#1}}
\newcommand{\labs}[2]{\slabs{#1}.\,{#2}}
\newcommand{\sslapp}{@}
\newcommand{\slapp}{\hspace*{1.5pt}}
\newcommand{\lapp}[2]{{#1}\slapp{#2}}

\newcommand{\avar}{x}

\newcommand{\snlvar}{\inMath{\mathsf{0}}}

\newcommand{\snlvarsucc}{\inMath{\mathsf{S}}}

\newcommand{\ater}{M}

\newcommand{\allter}{L}

\newcommand{\allteri}{\iap{\allter}}

\newcommand{\lambdahotg}{$\lambda$\nb-ho-\termdashgraph{}}
  \newcommand{\termdashgraph}{term-graph}
\newcommand{\lambdahotgs}{\lambdahotg{s}}

\newcommand{\alhotg}{{\cal G}}
\newcommand{\alhotgi}[1]{{\cal G}_{#1}}

\newcommand{\lambdaDFA}{$\lambda$\nb-DFA}
\newcommand{\lambdaDFAs}{\lambdaDFA{s}}

\newcommand{\backlink}{backlink}
\newcommand{\backlinks}{\backlink{s}}

\newcommand{\lambdatg}{$\lambda$\nb-\termdashgraph{}{}}
\newcommand{\lambdatgs}{\lambdatg{s}}

\newcommand{\altg}{G}
\newcommand{\altgi}{\iap{\altg}}

\newcommand{\unfsem}[1]{\llbracket{#1}\rrbracket_{\hspace*{-0.1pt}{\txtinflambdacal}}}
\newcommand{\sunfsem}{\unfsem\cdot}

\newcommand{\sunfsemeq}{{\iap{=}{\unfsem{\cdot}}}}
\newcommand{\unfsemeq}{\mathrel{\sunfsemeq}}
\newcommand{\snotunfsemeq}{{\iap{\neq}{\unfsem{\cdot}}}}
\newcommand{\notunfsemeq}{\mathrel{\snotunfsemeq}}

\newcommand{\graphsemC}[2]{\llbracket{#2}\rrbracket_{#1}}
\newcommand{\sgraphsemC}[1]{\graphsemC{#1}{\cdot}}

\newcommand{\classlhotgs}{{\cal H}}

\newcommand{\classltgs}{{\cal T}} 

\newcommand{\sreadback}{{\normalfont \textsf{rb}}}
\newcommand{\readback}{\funap{\sreadback}}

\newcommand{\scoll}{{{\text{\small\textbar}}\hspace{-0.73ex}\downarrow}}

\newcommand{\addvertnamedarkcyan}[2]{\path ({#1}) ++ (+0.05cm,-0.2cm) node[right] {$\scriptstyle{\darkcyan{#2}}$};}

\newcommand{\addvertprefixalert}[2]{\path ({#1}) ++ (-0.05cm,+0.2cm) node[left] {\textcolor{red}{$\scriptstyle{(#2)}$}};}

\newcommand{\addvertprefixposalert}[4]{\path ({#1}) ++ (-0.05cm,+0.2cm) ++ (#3,#4) node[left] {\textcolor{red}{$\scriptstyle{(#2)}$}};}

\newcommand{\LTS}{LTS}

\newcommand{\TSS}{TSS}

\newcommand{\abstractionprefix}{ab\-strac\-tion-pre\-fix}

\newcommand{\alphaconversion}{$\alpha$\nb-con\-ver\-sion}

\newcommand{\betareduction}{$\beta$\nb-re\-duc\-tion}

\newcommand{\combinatorreduction}{com\-bi\-na\-tor-re\-duc\-tion}

\newcommand{\graphrewriting}{graph-re\-writing}

\newcommand{\lambdacalculus}{$\lambda$\nb-calculus}

\newcommand{\lambdaterm}{$\lambda$\nb-term}
\newcommand{\lambdaterms}{\lambdaterm{s}}

\newcommand{\lambdaabstraction}{$\lambda$\nb-ab\-strac\-tion}
\newcommand{\lambdaabstractions}{\lambdaabstraction{s}}

\newcommand{\lambdaexpressions}{\lambdaexpressions}

\newcommand{\lambdaletrecterm}{$\txtlambdaletreccal$\nb-term}
\newcommand{\lambdaletrecterms}{\lambdaletrecterm{s}}
\newcommand{\lambdaletreccalterm}{$\lambdaletreccal$\nb-term}
\newcommand{\lambdaletreccalterms}{\lambdaletreccalterm{s}}

\newcommand{\lambdatermgraph}{$\lambda$\nb-term-graph}
\newcommand{\lambdatermgraphs}{\lambdatermgraph{s}}

\newcommand{\letfloating}{let-floa\-ting}

\newcommand{\maximalsharing}{max\-i\-mal-sha\-ring}

\newcommand{\languagesemantics}{lan\-guage-se\-man\-tics}
\newcommand{\processsemantics}{pro\-cess-se\-man\-tics}

\newcommand{\leftmostoutermost}{left\-most-out\-er\-most}

\newcommand{\LLEEpreserving}{LLEE-pre\-ser\-ving}

\newcommand{\loopbody}{loop-bo\-dy}
\newcommand{\loopentry}{loop-en\-try}
\newcommand{\loopelimination}{loop-eli\-mi\-na\-tion}

\newcommand{\multistep}{multi-step}
\newcommand{\multisteps}{\multistep{s}}

\newcommand{\nondeterministic}{non-de\-ter\-mi\-nistic}
\newcommand{\nonempty}{non-emp\-ty}

\newcommand{\finitestate}{fi\-nite-state}
\newcommand{\DFA}{DFA}

\newcommand{\NFA}{NFA}
\newcommand{\NFAs}{\NFA{s}}

\newcommand{\scopesharing}{scope-sha\-ring}
\newcommand{\scopepreserving}{scope-pre\-ser\-ving}
\newcommand{\contextsharing}{con\-text-sha\-ring}

\newcommand{\staticanalysis}{sta\-tic-anal\-y\-sis}
\newcommand{\structureconstrained}{struc\-ture-con\-strained}

\newcommand{\nonexample}{non-ex\-am\-ple}
\newcommand{\nonexamples}{\nonexample{s}}

\newcommand{\nontrivial}{non-triv\-i\-al}

\newcommand{\stexponefree}{$\stexpone$\nb-free}

\newcommand{\txtusof}{un\-der-star-\stexponefree}

\newcommand{\onetransition}{$\sone$\nb-tran\-si\-tion}
\newcommand{\onetransitions}{\onetransition{s}}

\newcommand{\nearcollapsed}{near-col\-lapsed}
\newcommand{\rightdistributivity}{right-distri\-bu\-ti\-vity}
\newcommand{\selfinverse}{self-in\-verse}
\newcommand{\superexponential}{super-ex\-po\-nen\-tial}
\newcommand{\scc}{scc}
\newcommand{\sccs}{\scc's}

\newcommand{\twincrystal}{twin-crys\-tal}
\newcommand{\twincrystals}{\twincrystal{s}}

\newcommand{\wellbehaved}{well-be\-haved}

\newcommand{\wellfounded}{well-foun\-ded}
\newcommand{\wellfoundedly}{\wellfounded{ly}}

\newcommand{\unfoldingsemantics}{un\-fol\-ding-se\-man\-tics}

\newcites{intro}{References} 
\newcites{lambdaletrec}{References} 
\newcites{procint}{References} 
\newcites{future}{References} 

\begin{document}
\maketitle

\begin{abstract}
  As a supplement to my talk at the workshop, 
    this extended abstract motivates and summarizes my work with co-authors
      on problems in two separate areas: 
        first, in the \lambdacalculus\ with \txtletrec,
           a universal model of computation, 
         and second, on Milner's process interpretation of regular expressions,  
           a proper subclass of the finite-state processes. 
  The aim of my talk was to motivate a transferal of ideas for workable concepts of \structureconstrained\ graphs: 
    from the problem of
      finding compact graph representations for terms in the \lambdacalculus\ with \txtletrec\
    to the problem of 
      recognizing finite process graphs that can be expressed by regular expressions. 
    In both cases the construction of \structureconstrained\ graphs 
      was expedient in order to enable to go back and forth easily 
       between, in the first case, \lambdaterms\ and term graphs, 
         and in the second case, regular expressions and process graphs.      
        
  The main focus here is on providing pointers 
    to my work with co-authors, in both areas separately.
  A secondary focus is on explaining directions of my present projects,
    and describing research questions of possibly general interest that have developed out of my work in these two areas.
\end{abstract}

\section{Introduction}%
  \label{intro}
%

The purpose of this extended abstract is to supplement my talk at the workshop 
                                                                               \cite{grab:2023:DCM} 
  with a brief description of my work with co-authors in two areas,
    including ample references.
While my work\-shop-pre\-sen\-ta\-tion covered similar topics as my talk \citeintro{grab:2018:TERMGRAPH} at TERMGRAPH~2018,
  and while the proceedings article \citeintro{grab:2019:EPTCS} for that workshop remains a useful resource,
    this article is a rewritten account with a detailed update on results that have been obtained in the meantime,
      and with an outlook on remaining challenging problems.

My talk \citeintro{grab:2023:DCM} at the workshop
  aimed at motivating
    a fruitful transferal of ideas between two areas on which I worked in the (a bit removed, and more recent) past:
      \lambdacalculus, and the implementation of functional programming languages (2009--2014),
        and the process theory of finite-state processes (\mbox{2005--6}, and from 2015).
My intention was to show,
  supported by many pictures:  
  How a solution to the problem of finding adequate graph representations for terms in the \lambdacalculus\ with \txtletrec, 
    a universal model of computation, 
  turned out to be very helpful in understanding process graphs that can be
    expressed by regular expressions (via Milner's process interpretation),
   a proper subclass of finite-state~processes. 
   
In both cases the definition of an adequate notion of structure-constrained (term or process) graph 
  was the key to solve a specific practical, and respectively, a theoretical problem. 
It was central that the structure-constrained graphs facilitate to go back and forth easily 
   between, on the one hand, terms in the \lambdacalculus\ with \txtletrec\ and term graphs, 
     and on the other hand, regular expressions and process graphs. 
The graph representations respect the appertaining operational semantics, but were conceived
with specific purposes in mind: to optimize functional programs in the
Lambda Calculus with \txtletrec ; and respectively, to reason with process graphs denoted
by regular expressions, and to decide recognizability of these graphs.
For a detailed comparison of the similarities and differences of
  the \structureconstrained\ graphs as defined in the  
    term graph semantics of terms in the \lambdacalculus\ with \txtletrec\ (see Section~\ref{lambdaletrec}),
      and the process (graph) semantics of regular expressions (see Section~\ref{procint}),
        we want to refer to Section~4 of \citeintro{grab:2019:EPTCS}.
          
Section~\ref{lambdaletrec} summarizes work by Jan Rochel and myself
  that led us to the definition, 
    and efficient implementation of maximal sharing for
      the higher-order terms in the \lambdacalculus\ with \txtletrec.
Specifically we formulated a representation-pipeline:
  Higher-order terms can be represented by, appropriately defined, 
    higher-order term graphs, then these can be encoded as first-order term graphs, 
      and subsequently those can in turn be represented as deterministic finite-state automata (DFAs).
Via these correspondences and DFA minimization, maximal shared forms of higher-order terms can be computed.
 
Section~\ref{procint} gives an overview of my work, in important parts done together with Wan Fokkink,
  on two \nontrivial\ problems that concern the process semantics of regular expressions.
In Milner's process semantics, regular expressions are interpreted as
  finite process graphs (or for that matter \nondeterministic\ \finitestate\ automata (NFAs)) 
    that are viewed as equal (as describing the same `behavior') if they are bisimilar. 
Unlike for the standard language interpretation,
not every finite process can be expressed, in this way, by a regular expression. 
This fact raised a \nontrivial\ recognition (or expressibility) problem, which was formulated
by Milner (1984) next to a completeness problem for an equational
proof system. 
In Section~\ref{procint} I report on the crucial steps that have led me to a solution of the completeness problem.

Finally Section~\ref{future} reports on my present projects, 
  and lists research questions that have developed out of my work in these two areas.

\bibliographystyleintro{eptcs}
\bibliographyintro{intro.bib}

\section{Compactifying Lambda-Letrec Terms}%
  \label{lambdaletrec}

This section gives an overview about work that Jan Rochel and I did in the framework of the NWO-project {Realizing Optimal Sharing (ROS)} at Utrecht University (2009--2014).%
    \footnote{This project was headed jointly by Vincent van Oostrom (rewriting and \lambdacalculus) and Doitse Swierstra (implementation of functional languages).
              The project was concluded successfully in June 2016 with Jan Rochel's defense of his thesis \citelambdaletrec{roch:2016}.}
It eventually led us to the definition and practical implementation of maximal sharing for terms in the \lambdacalculus\ with \txtletrec,
    the Core language for the compilation of functional programming languages.  

We started with the intention to study phenomena that arise practically for {optimal-sharing} implementations of the \lambdacalculus\ 
  (by graph-transformation schemes due to Lamping \citelambdaletrec{lamp:1989},
                                  and Kathail \citelambdaletrec{kath:1990}, and later interaction-net formalizations
    by Gonthier, Abadi, L\`{e}vy \citelambdaletrec{gont:abad:levy:1992}, and also van~Oostrom, van~de~Looij, Zwitserlood \citelambdaletrec{oost:looi:zwit:2004}),
    which are implementations of {optimal} or {parallel \betareduction} (due to L\`{e}vy \citelambdaletrec{levy:1978}).
For this purpose    
      Rochel wrote an impressive visualization and animation tool \citelambdaletrec{roch:2010} for transforming graphs by reducing graph-rewrite redexes per mouse-click.
        It produces beautifully rendered graphs that slowly float over the screen like bacteria in a liquid under a microscope.
This animation tool provided us with much room for experimentation.
We first tried to understand whether optimal implementations could render the so-called {static-argument transformation} unnecessary.
When we could not establish that, 
  we first tried to understand in how far the static-argument transformation
    changes the evaluation of programs with respect to usual scope-preserving graph evaluation.
As a consequence,     
  we partly turned our attention away from optimal evaluation 
    (in the hope that we would later come back to it with a better~understanding).

We started by generalizing the static-argument transformation to more general optimizations.    
      
\renewcommand{\descriptionlabel}[1]%
  {\hspace*{0ex}{\sf{#1}\hspace*{0ex}}}
  \begin{description}[leftmargin=*,itemindent=-1.25em]
    \item{} {\sf Parameter-dropping optimization transformations} 
      
      In 
         \citelambdaletrec{roch:grab:2011}
      we described an optimization transformation for the compilation of functional programs
      that drops parameters that are passed along unchanged between a number of recursive functions
      from the definitions of these functions.
      We used higher-order rewrite rules 
        to describe this generalization of the static-argument transformation  
        that permits the avoidance of repetitive evaluation patters \citelambdaletrec{roch:grab:2011}.
      We discovered later a close connection with Lambda Dropping due to Danvy and Schultz \citelambdaletrec{danv:schu:2000}.
 \end{description}
   
Realizing that we had moved on to terrain for which a strong theory had already been established, 
we set ourselves more ambitious goals:
First, to understand formally and conceptually
    the relationship between terms in the \lambdacalculus\ with \txtletrec\ ($\lambdaletreccal$) 
      and the infinite \lambdaterms\ they represent (in $\inflambdacal$, the infinitary \lambdacalculus).
Second, to find term graph representations of \lambdaletreccalterms\
  that are preserved under homomorphism (functional bisimularity). 
Finally third, we wanted to use possible answers for these two points
  to define maximally-shared representations of arbitrary \lambdaletreccalterms.  
Below we report on our results concerning these three goals.  

\begin{enumerate}[leftmargin=*,itemindent=-1.25em]
  \item{} {\sf Expressibility of infinite \lambdaterms\ by terms in $\lambdaletreccal$ (and in $\lambdamucal$).} 
    
    We studied the question: Which infinite \lambdaterms\ are (infinite) unfoldings of terms in \lambdaletreccal, the \lambdacalculus\ with \txtletrec,
      or (equivalent, but formally easier) in \lambdamucal, the \lambdacalculus\ with \txtmu? 
    Clearly, such infinite \lambdaterms\ have to be \emph{regular} in the sense 
      that their 
                 syntax-trees have only finitely many subtrees modulo \alphaconversion.
    However, while regularity is necessary for expressibility by a \lambdaletrecterm\ under infinite unfolding, it is not sufficient.
    What is missing is, intuitively, that the abstraction scopes in regular infinite \lambdaterms\ are not infinitely entangled. 
    We formulated this requirement in two different ways:
       that the infinite (regular) \lambdaterm\ in question 
      (i)~has only finitely many `generated subterms' 
            that are generated by a certain decomposition rewrite system that uses eager scope closure,
      (ii)~does not contain infinite `binding--capturing chains'.
    Both conditions delineate the \emph{strongly regular} infinite \lambdaterms\ among the regular ones. 
    For this concept we showed
      that an infinite \lambdaterm\ $\ater$ is the unfolding of a term in~\lambdaletreccal\ (resp.\ a term in~\lambdamucal)
        if and only if
      $\ater$ is strongly regular.
    For \lambdaletreccal-expressibility we showed that in \citelambdaletrec{grab:roch:2012:expressibility},
      and for \lambdamucal-expressibility in \citelambdaletrec{grab:roch:2013:RTA,grab:roch:2013:RTA:report};
        slides with many suggestive illustrations can be found in \citelambdaletrec{grab:2019:CLA}.
    
    Part of \citelambdaletrec{grab:roch:2012:expressibility}, and described separately in \citelambdaletrec{roch:grab:2013:IWC}, 
      is a non-trivial proof of confluence of a higher-order rewriting system that defines the unfolding semantics for \lambdaletreccalterms. 
    Furthermore in \citelambdaletrec{grab:roch:2013:IWC} we showed confluence of \letfloating\ operations on \lambdaletrecterms,
      obtaining a unique-normal-form result for \letfloating, 
        by using a higher-order rewriting system for the formalization of \letfloating.
                 
  \item{} {\sf Term graph representations of cyclic \lambdaterms.}
    \nopagebreak[4]\\[1ex]
    In \citelambdaletrec{grab:roch:2013:TERMGRAPH,grab:roch:2013:tgrfclt:ext:report}
    we systematically investigated a range of natural options for faithfully representing 
      the cyclic \lambdaterms\ in \lambdaletreccal\ 
    by higher-order term graphs (first-order term graphs with additional features that describe scopes),
      and by first-order term graphs (with specific~scope-delimiting~vertices). 
      
\begin{figure}[t!]
\begin{center}
\begin{tikzpicture}
    \pgfdeclarelayer{background}
    \pgfdeclarelayer{outerscope}
    \pgfdeclarelayer{innerscope}
    \pgfdeclarelayer{graph}
    \pgfdeclarelayer{boundingbox}
    \pgfdeclarelayer{caption}
    \pgfsetlayers{boundingbox,caption,background,outerscope,innerscope,graph}

\begin{pgfonlayer}{graph}
\matrix[anchor=north,row sep=0.35cm,column sep=0.4cm,every node/.style={circle,draw,thick,fill=white,scale=1,minimum size=0.4cm,inner sep=0pt}] at (1.5,0) { 
  & & \node(root-lhotg-3a){$\sslabs$}; \\
  & & \node(0-root-lhotg-3a){$\sslabs$}; \\ 
  & & \node(00-root-lhotg-3a){$\sslapp$}; \\
  & \node(000-root-lhotg-3a){$\sslapp$}; \\
  \node(0000-root-lhotg-3a){$\snlvar$}; & & \node(helper1)[fill=none,draw=none]{};\\   
  & & \node(helper2)[fill=none,draw=none]{}; & & & & & \node(001-root-lhotg-3a){$\snlvar$};\\
}; 
\draw[->,thick,chocolate,>=latex] ($(root-lhotg-3a.north) + (0pt,1pt) + (0pt,0.35cm)$) -- ($(root-lhotg-3a.north) + (0pt,1pt)$);
\draw[-,decoration={markings,mark=at position 1 with {\arrow[thick]{>}}},
        postaction={decorate}](root-lhotg-3a) to (0-root-lhotg-3a);
\draw[-,decoration={markings,mark=at position 1 with {\arrow[thick]{>}}},
        postaction={decorate}](0-root-lhotg-3a) to (00-root-lhotg-3a);
\draw[-,decoration={markings,mark=at position 1 with {\arrow[thick]{>}}},
        postaction={decorate}](00-root-lhotg-3a) to (000-root-lhotg-3a);
\draw[-,decoration={markings,mark=at position 1 with {\arrow[thick]{>}}},
        postaction={decorate}](000-root-lhotg-3a) to (0000-root-lhotg-3a);
\draw[-,decoration={markings,mark=at position 1 with {\arrow[thick]{>}}},
        postaction={decorate}](00-root-lhotg-3a) to (001-root-lhotg-3a);
\draw[->,decoration={markings,mark=at position 1 with {\arrow[thick]{>}}},
         postaction={decorate}](000-root-lhotg-3a) to[out=-50,in=50,distance=1.75cm] 
                  (00-root-lhotg-3a); 
\draw[-,decoration={markings,mark=at position 1 with {\arrow[thick]{>}}},
        postaction={decorate}](0000-root-lhotg-3a) -- ($ (helper1) + (2.65cm,0cm) $)  
                -- ($ (0-root-lhotg-3a) + (2.65cm,0cm) $)        
                -- (0-root-lhotg-3a);                  
\draw[-,decoration={markings,mark=at position 1 with {\arrow[thick]{>}}},
        postaction={decorate}](001-root-lhotg-3a)  -- ($ (helper2) + (3.2cm,0cm) $)
                -- ($ (root-lhotg-3a) + (3.2cm,0cm) $)
                -- (root-lhotg-3a);   
\end{pgfonlayer}

\begin{pgfonlayer}{innerscope}    
  \draw[draw=none,fill=red!27]
    (0000-root-lhotg-3a) -- ($ (helper1) + (2.65cm,0cm) $)  
           -- ($ (0-root-lhotg-3a)       + (2.65cm,0cm) $)        
           -- ($ (0-root-lhotg-3a)       + (-2.65cm,0cm) $)
           -- ($ (helper1) + (-2.65cm,0cm) $)
           -- (0000-root-lhotg-3a);           
\end{pgfonlayer}{innerscope}

\begin{pgfonlayer}{outerscope}                                 
  \draw[draw=none,fill=red!12]
    (001-root-lhotg-3a) -- ($ (helper2) + (3.2cm,0cm) $)
          -- ($ (root-lhotg-3a)    + (3.2cm,0cm) $)
          -- ($ (root-lhotg-3a)    + (-3.2cm,0cm) $)
          -- ($ (helper2) + (-3.2cm,0cm) $)
          -- (001-root-lhotg-3a);
\end{pgfonlayer}

\begin{pgfonlayer}{caption}
  \path[draw=none] (root-lhotg-3a) -- ($ (root-lhotg-3a) + (0cm,-4.7cm) $) node{\emph{$\lambda$-higher-order term graph}, version 1:};
  \path[draw=none] (root-lhotg-3a) -- ($ (root-lhotg-3a) + (0cm,-5.2cm) $) node{term graph {with \textcolor{red}{scope sets}}};
  \path[draw=none] (root-lhotg-3a) -- ($ (root-lhotg-3a) + (0cm,-5.6cm) $) node{{as structure constraints}};
\end{pgfonlayer}

\begin{pgfonlayer}{background}
\path[thick,dotted,
                   use as bounding box]
       ($ (root-lhotg-3a.north) + (-3.5cm,0.5cm) $)
    -- ($ (root-lhotg-3a.north) + (-3.5cm,-6.1cm) $)
    -- ($ (root-lhotg-3a.north) + (+3.5cm,-6.1cm) $)
    -- ($ (root-lhotg-3a.north) + (+3.5cm,0.5cm) $)
    -- ($ (root-lhotg-3a.north) + (-3.5cm,0.5cm) $);  
\end{pgfonlayer}{background}

\begin{pgfonlayer}{graph}
\matrix[anchor=north,row sep=0.35cm,column sep=0.4cm,every node/.style={circle,draw,thick,fill=white,scale=1,minimum size=0.4cm,inner sep=0pt}] at (9,0) 
                                                                                                                                                            {
  & & \node(root-lhotg-4a){$\sslabs$}; \\
  & & \node(0-lhotg-4a){$\sslabs$}; \\ 
  & & \node(00-lhotg-4a){$\sslapp$}; \\
  & \node(000-lhotg-4a){$\sslapp$}; \\
  \node(0000-lhotg-4a){$\snlvar$}; & & \node(helper1-lhotg-4a)[fill=none,draw=none]{};\\   
  & & \node(helper2)[fill=none,draw=none]{}; & & & & & \node(001-lhotg-4a){$\snlvar$};\\
}; 
\draw[->,thick,chocolate,>=latex] ($(root-lhotg-4a.north) + (0pt,1pt) + (0pt,0.35cm)$) -- ($(root-lhotg-4a.north) + (0pt,1pt)$);
\draw[-,decoration={markings,mark=at position 1 with {\arrow[thick]{>}}},
        postaction={decorate}](root-lhotg-4a) to (0-lhotg-4a);
\addvertnamedarkcyan{root-lhotg-4a}{\averti{0}}
\addvertprefixalert{root-lhotg-4a}{\niks}
\draw[-,decoration={markings,mark=at position 1 with {\arrow[thick]{>}}},
        postaction={decorate}](0-lhotg-4a) to (00-lhotg-4a);
\addvertnamedarkcyan{0-lhotg-4a}{\averti{1}}
\addvertprefixalert{0-lhotg-4a}{\averti{0}}
\draw[-,decoration={markings,mark=at position 1 with {\arrow[thick]{>}}},
        postaction={decorate}](00-lhotg-4a) to (000-lhotg-4a);
\addvertprefixalert{00-lhotg-4a}{\averti{0}\averti{1}}
\draw[-,decoration={markings,mark=at position 1 with {\arrow[thick]{>}}},
         postaction={decorate}](000-lhotg-4a) to (0000-lhotg-4a);
\addvertprefixalert{000-lhotg-4a}{\averti{0}\averti{1}}
\addvertprefixalert{0000-lhotg-4a}{\averti{0}\averti{1}}
\draw[-,decoration={markings,mark=at position 1 with {\arrow[thick]{>}}},
        postaction={decorate}](00-lhotg-4a) to (001-lhotg-4a);
\addvertprefixposalert{001-lhotg-4a}{\averti{0}}{-0.4ex}{-0.75ex}
\draw[-,decoration={markings,mark=at position 1 with {\arrow[thick]{>}}},
         postaction={decorate}](000-lhotg-4a) to[out=-50,in=50,distance=1.75cm] 
                  (00-lhotg-4a); 
\draw[-,decoration={markings,mark=at position 1 with {\arrow[thick]{>}}},
        postaction={decorate}](0000-lhotg-4a) -- ($ (helper1-lhotg-4a) + (2.65cm,0cm) $)  
                -- ($ (0-lhotg-4a) + (2.65cm,0cm) $)        
                -- (0-lhotg-4a);                  
\draw[-,decoration={markings,mark=at position 1 with {\arrow[thick]{>}}},
         postaction={decorate}](001-lhotg-4a)  -- ($ (helper2) + (3.2cm,0cm) $)
                -- ($ (root-lhotg-4a) + (3.2cm,0cm) $)
                -- (root-lhotg-4a);  
\end{pgfonlayer}                  
\begin{pgfonlayer}{innerscope}    
  \draw[draw=none,fill=none]
    (0000-lhotg-4a) -- ($ (helper1-lhotg-4a) + (2.65cm,0cm) $)  
           -- ($ (0-lhotg-4a)       + (2.65cm,0cm) $)        
           -- ($ (0-lhotg-4a)       + (-2.6cm,0cm) $)
           -- ($ (helper1-lhotg-4a) + (-2.6cm,0cm) $)
           -- (0000-lhotg-4a);           
\end{pgfonlayer}{innerscope}
\begin{pgfonlayer}{outerscope}                                 
  \draw[draw=none,fill=none]
    (001-lhotg-4a) -- ($ (helper2) + (3.2cm,0cm) $)
          -- ($ (root-lhotg-4a)    + (3.2cm,0cm) $)
          -- ($ (root-lhotg-4a)    + (-3.25cm,0cm) $)
          -- ($ (helper2) + (-3.25cm,0cm) $)
          -- (001-lhotg-4a);
\end{pgfonlayer}

\begin{pgfonlayer}{caption}
  \path[draw=none] (root-lhotg-3a) -- ($ (root-lhotg-4a) + (0cm,-4.7cm) $) node{\emph{$\lambda$-higher-order term graph}, version~2:};
  \path[draw=none] (root-lhotg-4a) -- ($ (root-lhotg-4a) + (0cm,-5.2cm) $) node{term graph {with \textcolor{red}{abstraction prefix function}}};
  \path[draw=none] (root-lhotg-4a) -- ($ (root-lhotg-4a) + (0cm,-5.6cm) $) node{{as structure constraint}};
\end{pgfonlayer}

\begin{pgfonlayer}{background}
  \path[thick,dotted,
                     use as bounding box]
         ($ (root-lhotg-4a.north) + (-3.5cm,0.5cm) $)
      -- ($ (root-lhotg-4a.north) + (-3.5cm,-6.1cm) $)
      -- ($ (root-lhotg-4a.north) + (+3.5cm,-6.1cm) $)
      -- ($ (root-lhotg-4a.north) + (+3.5cm,0.5cm) $)
      -- ($ (root-lhotg-4a.north) + (-3.5cm,0.5cm) $);  
\end{pgfonlayer}{background} 
  
\end{tikzpicture}  
\end{center}  
  \vspace*{-2ex}
\caption{\label{fig:lhotgs}%
         Translation of the \protect\lambdaletreccalterm\
           $\protect\allteri{0} 
              \protect\defdby 
            \protect\labs{\avar}{\labs{f}{\letin{r = \lapp{\lapp{f}{r}}{\avar}}{r}}}$ 
         into a $\protect\lambda$-higher-order term graph with scope sets \`{a} la Blom (left),
           and a $\protect\lambda$-h-o term graph $\graphsemC{\classlhotgs}{\allteri{0}}$ with an abstraction-prefix function (right). 
         Note that the inner scope has been chosen minimally here, applying eager scope closure. 
         (Non-eager scope \lambdahotgs\ can be defined as well, but are not expedient for maximal sharing.)
         } 
\end{figure}%
\begin{figure}[t!]
\begin{center}
\begin{tikzpicture}
    \pgfdeclarelayer{background}
    \pgfdeclarelayer{outerscope}
    \pgfdeclarelayer{innerscope}
    \pgfdeclarelayer{graph}
    \pgfdeclarelayer{boundingbox}
    \pgfdeclarelayer{caption}
    \pgfsetlayers{boundingbox,caption,background,outerscope,innerscope,graph}

\begin{pgfonlayer}{graph}
\matrix[anchor=north,row sep=0.35cm,column sep=0.4cm,every node/.style={circle,fill=white,draw,thick,scale=1,minimum size=0.4cm,inner sep=0pt}] at (-0.4,4.5) 
                                                                                                                                                           {
  & & \node(root-ltg-3){$\sslabs$}; \\
  & & \node(0-ltg-3){$\sslabs$}; \\ 
  & & \node(00-ltg-3){$\sslapp$}; \\
  & \node(000-ltg-3){$\sslapp$}; \\
  \node(0000-ltg-3){$\snlvar$}; & & \node(helper1-ltg-1)[draw=none,fill=none]{}; & & & \node(001-ltg-3)[draw,thick,circle,scale=1,minimum size=0.4cm,inner sep=0pt,red,fill=white]{${\snlvarsucc}$};\\   
  & & \node(helper2-ltg-1)[draw=none,fill=none]{}; & & & \node(0010-ltg-3){$\snlvar$};\\
};    
\draw[->,thick,chocolate,>=latex] ($(root-ltg-3.north) + (0pt,1pt) + (0pt,0.35cm)$) -- ($(root-ltg-3.north) + (0pt,1pt)$);
\draw[-,decoration={markings,mark=at position 1 with {\arrow[thick]{>}}},
        postaction={decorate}](root-ltg-3) to (0-ltg-3);
\draw[-,decoration={markings,mark=at position 1 with {\arrow[thick]{>}}},
        postaction={decorate}](0-ltg-3) to (00-ltg-3);
\draw[-,decoration={markings,mark=at position 1 with {\arrow[thick]{>}}},
        postaction={decorate}](00-ltg-3) to (000-ltg-3);
\draw[-,decoration={markings,mark=at position 1 with {\arrow[thick]{>}}},
        postaction={decorate}](000-ltg-3) to (0000-ltg-3);
\draw[-,decoration={markings,mark=at position 1 with {\arrow[thick]{>}}},
        postaction={decorate}](00-ltg-3) to (001-ltg-3);
\draw[-,decoration={markings,mark=at position 1 with {\arrow[thick]{>}}},
        postaction={decorate}](001-ltg-3) to (0010-ltg-3);
\draw[->,decoration={markings,mark=at position 1 with {\arrow[thick]{>}}},
         postaction={decorate}](000-ltg-3) to[out=-50,in=50,distance=1.75cm] 
                  (00-ltg-3); 
\draw[-](0000-ltg-3) -- ($ (001-ltg-3.west) + (-2pt,0pt) $);                  
\draw[-,red,thick,decoration={markings,mark=at position 1 with {\arrow[thick]{>}}},
              postaction={decorate}](001-ltg-3) -- ($ (helper1-ltg-1) + (2.65cm,0cm) $)  
               -- ($ (0-ltg-3) + (2.65cm,0cm) $)        
               -- (0-ltg-3);                  
\draw[-,decoration={markings,mark=at position 1 with {\arrow[thick]{>}}},
        postaction={decorate}](0010-ltg-3)  -- ($ (helper2-ltg-1) + (3.2cm,0cm) $)
                 -- ($ (root-ltg-3) + (3.2cm,0cm) $)
                 -- (root-ltg-3); 
\end{pgfonlayer}                  
\begin{pgfonlayer}{innerscope}    
  \draw[draw=none,fill=grey!27]
    (0000-ltg-3) -- ($ (helper1-ltg-1) + (2.65cm,0cm) $)  
           -- ($ (0-ltg-3)       + (2.65cm,0cm) $)        
           -- ($ (0-ltg-3)       + (-2.65cm,0cm) $)
           -- ($ (helper1-ltg-1) + (-2.65cm,0cm) $)
           -- (0000-ltg-3);           
\end{pgfonlayer}{innerscope}
\begin{pgfonlayer}{outerscope}                                 
  \draw[draw=none,fill=grey!12]
    (0010-ltg-3) -- ($ (helper2-ltg-1) + (3.2cm,0cm) $)
           -- ($ (root-ltg-3) + (3.2cm,0cm) $)
           -- ($ (root-ltg-3) + (-3.2cm,0cm) $)
           -- ($ (helper2-ltg-1) + (-3.2cm,0cm) $)
           -- (0010-ltg-3);
\end{pgfonlayer}

\begin{pgfonlayer}{caption}
  \path[draw=none] (root-ltg-3) -- ($ (root-ltg-3) + (0cm,-4.6cm) $) node{{first-order term graph with}};
  \path[draw=none] (root-ltg-3) -- ($ (root-ltg-3) + (0cm,-5.1cm) $) node{{\textcolor{red}{scope vertices with backlinks} (+ scope sets)}};
\end{pgfonlayer}

\begin{pgfonlayer}{boundingbox}                     
\path[thick,dotted,
                   use as bounding box]
       ($ (root-ltg-3.north) + (-3.5cm,0.5cm) $)
    -- ($ (root-ltg-3.north) + (-3.5cm,-5.6cm) $)
    -- ($ (root-ltg-3.north) + (+3.5cm,-5.6cm) $)
    -- ($ (root-ltg-3.north) + (+3.5cm,0.5cm) $)
    -- ($ (root-ltg-3.north) + (-3.5cm,0.5cm) $);
\end{pgfonlayer}

\begin{pgfonlayer}{graph}
\matrix[anchor=north,row sep=0.35cm,column sep=0.4cm,every node/.style={circle,draw,thick,scale=1,minimum size=0.4cm,inner sep=0pt}]  at (7.1,4.5) 
                                                                                                                                      {
  & & \node(root-ltg-2){$\sslabs$}; \\
  & & \node(0-ltg-2){$\sslabs$}; \\ 
  & & \node(00-ltg-2){$\sslapp$}; \\
  & \node(000-ltg-2){$\sslapp$}; \\
  \node(0000-ltg-2){$\snlvar$}; & & \node(helper1-ltg-2)[draw=none]{}; & & & \node(001-ltg-2){$\snlvarsucc$};\\   
  & & \node(helper2-ltg-2)[draw=none]{}; & & & \node(0010-ltg-2){$\snlvar$};\\
};    
\draw[->,thick,chocolate,>=latex] ($(root-ltg-2.north) + (0pt,1pt) + (0pt,0.35cm)$) -- ($(root-ltg-2.north) + (0pt,1pt)$);
\draw[-,decoration={markings,mark=at position 1 with {\arrow[thick]{>}}},
        postaction={decorate}](root-ltg-2) to (0-ltg-2);
\draw[-,decoration={markings,mark=at position 1 with {\arrow[thick]{>}}},
        postaction={decorate}](0-ltg-2) to (00-ltg-2);
\draw[-,decoration={markings,mark=at position 1 with {\arrow[thick]{>}}},
        postaction={decorate}](00-ltg-2) to (000-ltg-2);
\draw[-,decoration={markings,mark=at position 1 with {\arrow[thick]{>}}},
        postaction={decorate}](000-ltg-2) to (0000-ltg-2);
\draw[-,decoration={markings,mark=at position 1 with {\arrow[thick]{>}}},
        postaction={decorate}](00-ltg-2) to (001-ltg-2);
\draw[-,decoration={markings,mark=at position 1 with {\arrow[thick]{>}}},
        postaction={decorate}](001-ltg-2) to (0010-ltg-2);
\draw[->,decoration={markings,mark=at position 1 with {\arrow[thick]{>}}},
         postaction={decorate}](000-ltg-2) to[out=-50,in=50,distance=1.75cm] 
                  (00-ltg-2); 
\draw[-](0000-ltg-2) -- ($ (001-ltg-2.west) + (-2pt,0pt) $);                  
\draw[-,decoration={markings,mark=at position 1 with {\arrow[thick]{>}}},
        postaction={decorate}](001-ltg-2) -- ($ (helper1-ltg-2) + (2.65cm,0cm) $)  
               -- ($ (0-ltg-2) + (2.65cm,0cm) $)        
               -- (0-ltg-2);                  
\draw[-,decoration={markings,mark=at position 1 with {\arrow[thick]{>}}},
        postaction={decorate}](0010-ltg-2)  -- ($ (helper2-ltg-2) + (3.2cm,0cm) $)
                 -- ($ (root-ltg-2) + (3.2cm,0cm) $)
                 -- (root-ltg-2);                    
\end{pgfonlayer}

\begin{pgfonlayer}{caption}
  \path[draw=none] (root-ltg-2) -- ($ (root-ltg-2) + (0cm,-4.6cm) $) node{\emph{$\lambda$\nb-term graph}: first-order term graph};
  \path[draw=none] (root-ltg-2) -- ($ (root-ltg-2) + (0cm,-5.1cm) $) node{with scope vertices with backlinks};
\end{pgfonlayer}

\begin{pgfonlayer}{boundingbox}                     
\path[thick,dotted,
                   use as bounding box]
       ($ (root-ltg-2.north) + (-3.5cm,0.5cm) $)
    -- ($ (root-ltg-2.north) + (-3.5cm,-5.6cm) $)
    -- ($ (root-ltg-2.north) + (+3.5cm,-5.6cm) $)
    -- ($ (root-ltg-2.north) + (+3.5cm,0.5cm) $)
    -- ($ (root-ltg-2.north) + (-3.5cm,0.5cm) $);
\end{pgfonlayer}

\end{tikzpicture}  
\end{center}  
  \vspace*{-2ex}
\caption{\label{fig:ltgs}%
         Translation of the \protect\lambdaletreccalterm\
           $\protect\allteri{0} 
              \protect\defdby 
            \protect\labs{\avar}{\labs{f}{\letin{r = \lapp{\lapp{f}{r}}{\avar}}{r}}}$ 
         into a $\protect\lambda$-term-graph $\graphsemC{\classltgs}{\allteri{0}}$
           by adding a scope vertex delimiting the inner scope 
             to the $\protect\lambda$-higher-order term graphs in Fig.~\protect\ref{fig:lhotgs},
               and by then dropping the scope sets (which now can be reconstructed as well as a corresponding abstraction prefix function). 
         While the backlink from the left variable vertex to its binding abstraction vertex
          is drawn suggestively along the scope border, it does not target the scope-delimiting vertex, 
            but continues invisibly below the backlink of that scope-delimiting vertex onwards to the commonly targeted abstraction vertex.
         (While not relevant for maximal sharing,
            relaxing the condition of eager scope closure for \protect\lambdahotgs\
              can be dealt with by an adapted encoding as first-order term graph.)    
         }
\end{figure}%
     
    As the result of this analysis we arrived at a natural class of higher-order term graphs
      (of `\lambdahotgs', see below) 
        that can be implemented faithfully as first-order term graphs (`\lambdatgs', see below).
    The basis of the higher-order term graphs (as well as of their first-order implementations) for representing \lambdaletreccalterms\
      are first-order term graphs with three different kinds of vertex labels:
      \begin{itemize}[itemsep=0.5ex]
        \item
          unary symbols $\sslabs$ for abstraction vertices,
        \item
          binary symbols $\sslapp$ for application vertices, and
        \item 
          unary symbols $\snlvar$ for nameless variable vertices that enable \backlinks\ to the binding abstraction vertices. 
      \end{itemize}     
    The first-order \lambdatgs\ also permit:
    \begin{itemize}
      \item
         binary symbols $\snlvarsucc$ for scope-delimiting vertices that facilitate backlinks to the abstraction vertices whose scope they close.   
    \end{itemize}       
       
    With this preparations we can now explain the higher-order \lambdahotgs\ and first-order \lambdatgs\
        in more detail. For the precise definitions and statements we refer to~%
          \citelambdaletrec{grab:roch:2013:TERMGRAPH,grab:roch:2013:tgrfclt:ext:report,grab:roch:2014:ICFP}.
    
    \begin{enumerate}[label={(\roman{*})},leftmargin=*,align=left,itemindent=0em,itemsep=0.5ex]
      \item[\textsf{\lambdahotgs}] appear in two versions:
          
        \begin{enumerate}[label={(\roman{*})},leftmargin=0.2em,align=left,itemsep=0.25ex]
          \item[\textsf{\lambdahotgs\ with scope-sets}]   
            are extensions of first-order term graphs with vertex labels $\sslabs$, $\sslapp$, and $\snlvar$
              by adding, to each abstraction vertex $\bvert$, a scope set that consists of all vertices in the scope of $\bvert$.
            The scope sets of abstraction vertices in a \lambdahotg\ satisfy a number of conditions that safeguard
              that (i)~scopes are nested, (ii)~scopes arise by eager scope closure,
                and (iii)~each variable vertex is contained in the scope of the abstraction vertex to which its backlink points to.
            In this way, scope sets aggregate scope information that is available locally at the abstraction vertices.
                
            \lambdatermgraphs\ with scope sets are an adaptation of Blom's of higher-order term graphs
              with scope~sets~\citelambdaletrec{blom:2001} to representing the cyclic \lambdaterms\ in \lambdaletreccal\ 
                (and the strongly regular infinite \lambdaterms\ in \inflambdacal).  
            For an example, see Figure~\ref{fig:lhotgs} on the left for the translation of a (variant) fixed-point combinator
              into a \lambdahotg\ with (eager-scope) scope sets. 
            
          \item[\textsf{\lambdahotgs\ with abstraction-prefix function}] 
            are extensions of first-order term graphs with vertex labels $\sslabs$, $\sslapp$, and $\snlvar$
              by adding an abstraction prefix function:
            That function assigns, to each vertex $\bvert$,
                  an abstraction prefix $(\averti{1}\ldots\averti{n})$ consisting of a word of abstraction vertices that lists those abstractions
                    (from the top down)
                    for which $\bvert$ is in their `extended scope' (transitive closure of scope relation) as obtained by eager scope closure.
            Abstraction prefixes aggregate scope information that then is locally available at individual vertices.
            
            See Figure~\ref{fig:lhotgs} on the right for the translation of a (variant) fixed-point combinator
              into a \lambdahotg\ with abstraction prefixes (obtained by eager scope closure).
        \end{enumerate}     
        
        \noindent
        In both versions of \lambdahotg,
          the added constraints guarantee
            that each variable vertex (with label $\snlvar$)
              has a backlink to the binding $\lambda$-abstraction vertex.
        A bijective correspondence can be shown to exist between both versions of \lambdahotgs\ 
          (see \citelambdaletrec{grab:roch:2013:TERMGRAPH,grab:roch:2013:tgrfclt:ext:report}).
        
      \item[\textsf{\lambdatgs}] \mbox{}
        are first-order term graphs that represent \lambdahotgs\ of both kinds as above.
          Scopes are delimited, again by using eager scope closure,
            by scope-delimiting vertices (with label $\snlvarsucc$) that have backlinks to the abstraction vertex whose scope they declare closed.
            Variable vertices (with label $\snlvar$) have backlinks to the binding $\lambda$-abstraction vertex.
            
        See Figure~\ref{fig:ltgs} for the encoding of the \lambdahotgs\ in Figure~\ref{fig:lhotgs}
          into a \lambdatg. For this purpose a scope-delimiter vertex with label $\snlvarsucc$ 
            is used to represent the (eager) closure of the inner scope.
    \end{enumerate}          
          
    The conditions underlying \lambdahotgs\ and \lambdatgs\ (see \citelambdaletrec{grab:roch:2013:TERMGRAPH,grab:roch:2013:tgrfclt:ext:report})
      guarantee that they represent finite or infinite closed \lambdaterms; that is, they do not contain meaningless parts.      
    Both \lambdahotgs\ (all two versions) and \lambdatgs\ induce appropriate concepts of homomorphism (functional bisimulation) and bisimulation. 
    Homomorphisms increase sharing, and introduce a sharing (partial) order.
    Bisimulations preserve the unfolding semantics (as do homomorphisms).
    We established in \citelambdaletrec{grab:roch:2013:TERMGRAPH,grab:roch:2013:tgrfclt:ext:report} 
      a bijective correspondence between \lambdahotgs\ and \lambdatgs\ that preserves and reflects homomorphisms,
        and hence the sharing (partial) order.
    These results form the basis of the maximal-sharing method, see below.
            
    The property that is of the most central importance for the \maximalsharing\ method 
      is that homomorphisms (functional bisimulations) between first-order term graphs
        preserve \lambdatgs: if $\agraphi{1}$ is a \lambdatg, and $\agraphi{1} \funbisim \agraphi{2}$ for a term graph $\agraphi{2}$
          (there is a homomorphism from $\agraphi{1}$ to $\agraphi{2}$), then also $\agraphi{2}$ is a \lambdatg. 
    For this property to hold, eager scope closure is crucial.%
      \footnote{%
        While that is not relevant for the maximal-sharing method (for which use of eager scope closure is essential),
          we mention as an aside that this restriction can be circumnavigated:
            a generalization of the preservation property can also be shown for a different kind of encoding
            of (also non-eager-scope) \lambdahotgs\ into first-order term graphs
              (see Remark 7.10 in \cite{grab:roch:2013:tgrfclt:ext:report}).}

\input{figs/fig-maxsharing-example.tex}

  \item{} {\sf Maximal sharing in \lambdaletreccal.} 
    
    For defining maximally shared versions of terms in \lambdaletreccal\ in a natural way
      we defined a `representation pipeline' in \citelambdaletrec{grab:roch:2014:ICFP,grab:roch:2014:maxsharing-arxiv}
        (see Figure~\ref{fig:maxsharing:method} for a suggestive illustration):
    First we linked \lambdaletreccalterms\ by an interpretation function $\graphsemC{\classlhotgs}{\cdot}$
      to the class $\classlhotgs$ of \lambdahotgs\ that we formulated earlier in \citelambdaletrec{grab:roch:2013:TERMGRAPH,grab:roch:2013:tgrfclt:ext:report}.
    Then we extended $\graphsemC{\classlhotgs}{\cdot}$ by using the representation 
          of \lambdahotgs\ as \lambdatgs\ (first-order term graphs) from \citelambdaletrec{grab:roch:2013:TERMGRAPH,grab:roch:2013:tgrfclt:ext:report}
            to define an interpretation function $\graphsemC{\classltgs}{\cdot}$ of \lambdaletreccalterms\ to the class $\classltgs$ of \lambdatgs.
    For this representation pipeline we showed that
      unfolding equivalence $\sunfsemeq$ of \lambdaletreccalterms\ is faithfully represented by bisimulation equivalence~$\sbisim$
        on \lambdahotgs, and equally, by bisimulation equivalence~$\sbisim$ on \lambdatgs.
    
    Then we defined a readback operation $\sreadback$ on \lambdatgs\ in the class $\classltgs$ (see also in Figure~\ref{fig:maxsharing:method})
      with the property that the interpretation operation $\graphsemC{\classltgs}{\cdot}$ is a left-inverse of $\sreadback$ on $\classltgs\,$: 
      \begin{equation*}
        \graphsemC{\classltgs}{\cdot} \:\circ\: \sreadback = \sidfunon{\classltgs} \quad \text{(modulo isomorphism).}
      \end{equation*}  
      
    These three operations facilitate to compute, for any given \lambdaletreccalterm~$\allter$,
      a maximally shared form $\allteri{0}$, by the following three-step procedure (see Figure~\ref{fig:maxsharing:method}):
      \begin{enumerate}[label={(\roman{*})},leftmargin=*,align=left]
        \item[(\textsf{interpret})]
          from $\allter$ its interpretation $\graphsemC{\classltgs}{\allter}$ as \lambdatg\ is obtained,
        \item[(\textsf{collapse})]
          from the \lambdatg\ $\graphsemC{\classltgs}{\allter}$ its bisimulation collapse $\agraphi{0}$ is computed,
            which is again a \lambdatg\ in $\classltgs$ (due to preservation of \lambdatgs\ along functional bisimulations),
        \item[(\textsf{readback})]
          from the collapsed \lambdatg\ $\agraphi{0}$ its readback $\readback{\agraphi{0}}$ is computed,
            thereby obtaining the term $\allteri{0} \defdby \readback{\agraphi{0}}$ as a maximally shared form $\allter$
              with $\allteri{0} \unfsemeq \allter$ (and hence so that $\allteri{0}$ has the same infinite unfolding as $\allter$).
      \end{enumerate}  
    This procedure permits an efficient implementation.
      We could derive its complexity as (at about) quadratic in the size of the input \lambdaletrecterm.%
\begin{figure}[t!]
\begin{center}
\begin{tikzpicture}[scale=1]
\matrix[row sep=0.35cm,column sep=0.4cm,every node/.style={circle,draw,thick,scale=1,minimum size=0.4cm,inner sep=0pt}] at (4,0) {
  & & \node(g1-root){$\sslabs$}; \\
  & & \node(g1-0){$\sslabs$};    \\ 
  & & \node(g1-00){$\sslapp$};   \\
  &   \node(g1-000){$\sslapp$};  \\
  \node(g1-0000){$\snlvar$}; 
      & & \node(g1-helper1)[draw=none]{}; 
    & & & \node(g1-001){$\snlvarsucc$}; \\   
      & & \node(g1-helper2)[draw=none]{}; 
  & & & \node(g1-0010){$\snlvar$};\\
};    
\draw[->,thick,chocolate,>=latex] ($(g1-root.north) + (0pt,1pt) + (0pt,0.305cm)$) -- ($(g1-root.north) + (0pt,1pt)$);
\draw[-,decoration={markings,mark=at position 1 with {\arrow[thick]{>}}},
        postaction={decorate}](g1-root) to (g1-0);
\draw[-,decoration={markings,mark=at position 1 with {\arrow[thick]{>}}},
        postaction={decorate}](g1-0) to (g1-00);
\draw[-,decoration={markings,mark=at position 1 with {\arrow[thick]{>}}},
        postaction={decorate}](g1-00) to (g1-000);
\draw[-,decoration={markings,mark=at position 1 with {\arrow[thick]{>}}},
        postaction={decorate}](g1-000) to (g1-0000);
\draw[-,decoration={markings,mark=at position 1 with {\arrow[thick]{>}}},
        postaction={decorate}](g1-00) to (g1-001);
\draw[-,decoration={markings,mark=at position 1 with {\arrow[thick]{>}}},
        postaction={decorate}](g1-001) to (g1-0010);
\draw[-,decoration={markings,mark=at position 0.999 with {\arrow[thick]{>}}},
        postaction={decorate}](g1-000) to[out=-45,in=43,distance=1.75cm] 
                  (g1-00); 
\draw[-](g1-0000) -- ($ (g1-001.west) + (-2pt,0pt) $);                  
\draw[-,decoration={markings,mark=at position 1 with {\arrow[thick]{>}}},
        postaction={decorate}](g1-001) -- ($ (g1-helper1) + (2.65cm,0cm) $)  
               -- ($ (g1-0) + (2.65cm,0cm) $)        
               -- (g1-0);                  
\draw[-,decoration={markings,mark=at position 1 with {\arrow[thick]{>}}},
        postaction={decorate}](g1-0010)  -- ($ (g1-helper2) + (3.2cm,0cm) $)
                 -- ($ (g1-root) + (3.2cm,0cm) $)
                 -- (g1-root);

\matrix[row sep=0.35cm,column sep=0.4cm,every node/.style={circle,draw,thick,scale=1,minimum size=0.4cm,inner sep=0pt}] at (-4,0){
  & & & \node(g2-root){$\sslabs$}; \\ 
  & & & \node(g2-0){$\sslabs$}; \\     
  & & & \node(g2-00){$\sslapp$}; \\
  & &   \node(g2-000){$\sslapp$}; & &  \\
  & & & \node(g2-0001){$\sslapp$}; & \\
  & &   \node(g2-00010){$\sslapp$}; & & \\
        \node(g2-0000){$\snlvar$}; 
    &   \node(g2-000100){$\snlvar$}; & & \node(g2-helper1)[draw=none]{}; 
    & & & \node(g2-00011){$\snlvarsucc$};
        & \node(g2-001){$\snlvarsucc$}; \\ 
  & & & \node(g2-helper2)[draw=none]{}; 
    & & & \node (g2-000110){$\snlvar$};
        & \node(g2-0010){$\snlvar$}; \\   
};    
\draw[->,thick,chocolate,>=latex] ($(g2-root.north) + (0pt,1pt) + (0pt,0.305cm)$) -- ($(g2-root.north) + (0pt,1pt)$);
\draw[-,decoration={markings,mark=at position 1 with {\arrow[thick]{>}}},
        postaction={decorate}](g2-root) to (g2-0);
\draw[-,decoration={markings,mark=at position 1 with {\arrow[thick]{>}}},
        postaction={decorate}](g2-0) to (g2-00);
\draw[-,decoration={markings,mark=at position 1 with {\arrow[thick]{>}}},
        postaction={decorate}](g2-00) to (g2-000);
\draw[-,decoration={markings,mark=at position 1 with {\arrow[thick]{>}}},
        postaction={decorate}](g2-000) to (g2-0000);
\draw[-,decoration={markings,mark=at position 1 with {\arrow[thick]{>}}},
        postaction={decorate}](g2-00) to (g2-001);
\draw[-,decoration={markings,mark=at position 0.999 with {\arrow[thick]{>}}},
        postaction={decorate}](g2-00010) to[out=-45,in=43,distance=1.75cm] (g2-00); 
\draw[-,decoration={markings,mark=at position 1 with {\arrow[thick]{>}}},
        postaction={decorate}](g2-000) to (g2-0001);
\draw[-,decoration={markings,mark=at position 1 with {\arrow[thick]{>}}},
        postaction={decorate}](g2-0001) to (g2-00011);
\draw[-,decoration={markings,mark=at position 1 with {\arrow[thick]{>}}},
         postaction={decorate}](g2-0001) to (g2-00010);
\draw[-,decoration={markings,mark=at position 1 with {\arrow[thick]{>}}},
        postaction={decorate}](g2-00010) to (g2-000100);
\draw[-,decoration={markings,mark=at position 1 with {\arrow[thick]{>}}},
         postaction={decorate}](g2-00011) to (g2-000110);
\draw[-,decoration={markings,mark=at position 1 with {\arrow[thick]{>}}},
         postaction={decorate}](g2-001) to (g2-0010);
\draw[-,decoration={markings,mark=at position 1 with {\arrow[thick]{>}}},
        postaction={decorate}](g2-001) -- ($ (g2-helper1) + (3.2cm,0cm) $) 
               -- ($ (g2-0)       + (3.2cm,0cm) $)
               -- (g2-0);
\draw[-](g2-00011) --  ($ (g2-001.west) + (-2pt,0pt) $);               
\draw[-](g2-0000) --  ($ (g2-000100.west) + (-2pt,0pt) $); 
\draw[-](g2-000100) --  ($ (g2-00011.west) + (-2pt,0pt) $);                             
\draw[-](g2-0010) -- ($ (g2-helper2) + (3.7cm,0cm) $)
                  -- ($ (g2-root) +    (3.7cm,0cm) $)
                  -- (g2-root);
\draw[-](g2-000110) -- ($ (g2-0010.west) + (-2pt,0pt) $);

\draw[dashed,magenta,bend right,distance=2.5cm] (g1-root) to (g2-root);
\draw[dashed,magenta,bend right,distance=2.25cm] (g1-0) to (g2-0);
\draw[dashed,magenta,bend right,distance=2cm] (g1-00) to (g2-00);  
\draw[dashed,magenta,bend right,distance=1.5cm] (g1-000) to (g2-000);
\draw[dashed,magenta,bend right,distance=1.75cm] (g1-001) to (g2-001);
\draw[dashed,magenta,out=225,in=-30,distance=4.5cm] (g1-0000) to (g2-0000);  
\draw[dashed,magenta,out=220,in=-30,distance=4.cm] (g1-0000) to (g2-000100);
\draw[dashed,magenta] (g1-00) to (g2-0001);
\draw[dashed,magenta] (g1-000) to (g2-00010);
\draw[dashed,magenta,bend right,distance=2cm] (g1-001) to (g2-00011);
\draw[dashed,magenta,bend left,distance=3cm] (g1-0010) to (g2-000110);
\draw[dashed,magenta,bend left,distance=2cm] (g1-0010) to (g2-0010);

\matrix[row sep=0.35cm,column sep=0.4cm] at (0.9,1) {    
  \node (bisimsym) {{\Large $\magenta{\sfunbisim}$}};
  \\
};

\path (3.75,-4.6) ++ (0cm,0cm) node{$\graphsemC{\classltgs}
                                                {
  \underbrace{\labs{\avar}{\labs{f}{\letin{r = \lapp{\lapp{f}{r}}{\avar}}{r}}}}_{\allteri{0}}                                                 
                                                  }$};

\path (-3.75,-4.6) ++ (0cm,0cm) node{$\graphsemC{\classltgs}
                                               {
  \underbrace{\labs{\avar}{\labs{f}{\letin{r = \lapp{\lapp{f}{(\lapp{\lapp{f}{r}}{\avar})}}{\avar}}{r}}}}_{\allter}
                                                }$};

\end{tikzpicture}
\end{center}  
  \vspace*{-2ex}
  \caption{\label{fig:ltg:L:funbisim:ltg:L0}%
           Compactification of the \protect\lambdaletreccalterm~$\allter$, 
             a redundant form of a variant fixed-point combinator (compare with the forms in Figure~\ref{fig:maxsharing:method}),
               to the more compact \protect\lambdaletreccalterm~$\allteri{0}$.
           The $\protect\lambda$\protect\nb-term-graph interpretations $\protect\graphsemC{\protect\classltgs}{\protect\allter}$ 
                                                                   and $\protect\graphsemC{\protect\classltgs}{\protect\allteri{0}}$
             of the \protect\lambdaletreccal\ terms $\protect\allter$ and $\protect\allteri{0}$ 
           are bisimilar.
           Indeed the links $\hspace*{-0.75em}\protect\picbisimlink$ form a functional bisimulation $\magenta{\protect\sfunbisim}$ from $\protect\graphsemC{\protect\classltgs}{\protect\allter}$
                                                                                                  to $\protect\graphsemC{\protect\classltgs}{\protect\allteri{0}}$,
           of which the \protect\lambdatermgraph\ $\protect\graphsemC{\protect\classltgs}{\protect\allteri{0}}$ is in bisimulation-collapsed form.                                                   
           }
\end{figure}
    See Figure~\ref{fig:ltg:L:funbisim:ltg:L0} for an example of the collapse step on the \lambdatg\ interpretation $\graphsemC{\classltgs}{\allter}$
      of an inefficient version $\allter$ of a (slight variation of a) fixed-point combinator to obtain the \lambdatg\ interpretation $\graphsemC{\classlhotgs}{\allteri{0}}$   
        that obtains a more efficient and compact version $\allteri{0}$ of such a combinator.
    
    A straightforward adaptation of this procedure permits to obtain also an efficient algorithm
      for deciding \unfoldingsemantics\ equality $\sunfsemeq$ of any two given \lambdaletreccalterms~$\allteri{1}$ and $\allteri{2}$
        by the following two-step procedure:
      \begin{enumerate}[label={(\roman{*})},leftmargin=*,align=left]
        \item[(\textsf{interpret})]
          obtain the \lambdatg\ interpretations $\agraphi{1} \defdby \graphsemC{\classltgs}{\allteri{1}}$ of $\allteri{1}$
            and $\agraphi{2} \defdby \graphsemC{\classltgs}{\allteri{2}}$ of $\allteri{2}$;
        \item[(\textsf{check-bisim})]
          check bisimilarity of $\agraphi{1}$ and $\agraphi{2}$;
            if $\agraphi{1} \bisim \agraphi{2}$ holds, conclude that $\allteri{1} \unfsemeq \allteri{2}$ holds
              (that is, $\allteri{1}$ and $\allteri{2}$ have the same infinite unfolding),
                otherwise $\allteri{1} \notunfsemeq \allteri{2}$ holds. 
      \end{enumerate}  
    
    We implemented both the \maximalsharing\ method and the decision procedure for unfolding equivalence $\sunfsemeq$
      by a prototype implementation \citelambdaletrec{roch:grab:2014:maxsharing:tool} that is available on Haskell's Hackage platform.     
    For the efficient implementation of these methods we extended the representation pipeline
      from \lambdatgs\ further to \lambdaDFAs, by which we mean representations of \lambdaletreccalterms\ as deterministic \finitestate\ automata. 
    In this way, unfolding equivalence $\sunfsemeq$ of \lambdaletreccalterms\
      is represented as language equivalence of \lambdaDFAs,
        and so we could use for the implementation \citelambdaletrec{roch:grab:2014:maxsharing:tool} 
          that bisimulation collapse of \lambdatgs\ is faithfully represented by state minimization of \lambdaDFAs.
\end{enumerate}

At the end of this section I want to mention a concept that Vincent van Oostrom
  suggested after seeing the concept of \lambdatgs\ in Jan Rochel's thesis \citelambdaletrec{roch:2016}:
    the concept of `nested term graphs'.

\renewcommand{\descriptionlabel}[1]%
  {\hspace*{0ex}{\sf{#1}\hspace*{0ex}}}
  \begin{description}[leftmargin=*,itemindent=-1.25em]
    \item{} {\sf Nested Term Graphs} 
        
      Motivated by the results on term graph representations and maximal sharing for \lambdaletrecterms,
      Vincent van Oostrom and I formulated a concept of nested term graph \citelambdaletrec{grab:oost:2015}.
      Instead of describing scopes by additional features like scope sets or an \abstractionprefix\ function
        in order to define constraints that guarantee that scopes are nested,
          we introduced `nesting' itself as a structuring concept. 
      This means that we permitted nesting of first-order term graphs into vertices of other first-order term graphs.
        In this manner, \wellfoundedly\ nested first-order term graphs can be defined by induction.     
      We studied the behavioral semantics of nested term graphs in~\citelambdaletrec{grab:oost:2015},
        and also showed, in analogy with the faithful encoding of \lambdahotgs\ as \lambdatgs, 
          that nested term graphs can be encoded by first-order term graphs
            faithfully (in the sense of preserving the respective unfolding semantics).
        
      Nested term graphs not only provide a natural formalization the maximal-sharing method developed in \citelambdaletrec{grab:roch:2013:TERMGRAPH,grab:roch:2014:ICFP},
      but they make it much more broadly applicable, also outside of Lambda Calculus.
  \end{description}

\bibliographystylelambdaletrec{eptcs}
\bibliographylambdaletrec{lambdaletrec.bib}

\section{Proving Bisimilarity between Regular-Expression Processes}%
  \label{procint}

This section motivates, summarizes, and provides references to my work on Milner's process semantics of regular expressions \citeprocint{miln:1984}.
  An important part of it (leading to \citeprocint{grab:fokk:2020:lics,grab:fokk:2020:lics:arxiv})
    was done in close collaboration (2015--2020) 
                                    with Wan Fokkink who had stimulated me to work on Milner's question already in 2005.
    While this section focuses on my work on Milner's axiomatization questions (see \ref{A} below),
      my current work on the expressibility question (see \ref{E} below) will be mentioned in Section~\ref{future}. 

Milner introduced a process semantics $\procsem{\cdot}$ in \citeprocint{miln:1984} for regular expressions (conceived by Kleene \citeprocint{klee:1951})
  that refines the standard language semantics $\langsem{\cdot}$ (defined by Copi, Elgot, Wright \citeprocint{copi:elgot:wrig:1958}). 
For regular expressions $\astexp$ 
  that are constructed from constants $\stexpzero$, $\stexpone$, letters over a given set $\actions$
    with the binary operators $\sstexpsum$ and $\sstexpprod$, and the unary operator $\stexpit{(\cdot)}$,
Milner first defined a process interpretation $\procint{\astexp}$ 
  that can informally be described as follows:
    $\stexpzero$ is interpreted as a deadlocking process without any observable behavior,
      $\stexpone$ as a process that terminates successfully immediately,
        letters from the set $\actions$ stand for atomic actions that lead to successful termination;
    the binary operators $\sstexpsum$ and $\sstexpprod$ are interpreted as the operations of choice and concatenation of two processes, respectively,
      and the unary star operator $\stexpit{(\cdot)}$ is interpreted as the operation of unbounded iteration of a process, 
        but with the option to terminate successfully before~each~iteration.  
        
Milner formalized this process interpretation in \citeprocint{miln:1984} as process graphs
  that are defined by induction on the structure of regular expressions.   
But soon afterwards a formal definition by means of a transition system specification (\TSS)
  that defines a labeled transition system (\LTS) became more common.
%
\begin{figure}[tp]
\begin{gather*}
 \begin{aligned}
   &
   \AxiomC{\phantom{$\terminates{\stexpone}$}}
   \UnaryInfC{$\terminates{\stexpone}$}
   \DisplayProof
   & \hspace*{-1.5ex} &
   \AxiomC{$ \terminates{\astexpi{1}} $}
   \UnaryInfC{$ \terminates{(\stexpsum{\astexpi{1}}{\astexpi{2}})} $}
   & \hspace*{2ex} &
   \AxiomC{$ \terminates{\astexpi{i}} $}
   \RightLabel{\small $(i\in\setexp{1,2})$}
   \UnaryInfC{$ \terminates{(\stexpsum{\astexpi{1}}{\astexpi{2}})} $}
   \DisplayProof
   & \hspace*{2ex} &
   \AxiomC{$\terminates{\astexpi{1}}$}
   \AxiomC{$\terminates{\astexpi{2}}$}
   \BinaryInfC{$\terminates{(\stexpprod{\astexpi{1}}{\astexpi{2}})}$}
   \DisplayProof
   & \hspace*{2ex} &
   \AxiomC{$\phantom{\terminates{\stexpit{\astexp}}}$}
   \UnaryInfC{$\terminates{(\stexpit{\astexp})}$}
   \DisplayProof
 \end{aligned} 
 \\[1ex]
 \begin{aligned}
   & 
   \AxiomC{$\phantom{a \:\lt{a}\: \stexpone}$}
   \UnaryInfC{$a \:\lt{a}\: \stexpone$}
   \DisplayProof
   & &
   \AxiomC{$ \astexpi{i} \:\lt{a}\: \astexpacci{i} $}
   \RightLabel{\small $(i\in\setexp{1,2})$}
   \UnaryInfC{$ \stexpsum{\astexpi{1}}{\astexpi{2}} \:\lt{a}\: \astexpacci{i} $}
   \DisplayProof 
   & &
   \AxiomC{$ \astexpi{1} \:\lt{a}\: \astexpacci{1} $}
   \UnaryInfC{$ \stexpprod{\astexpi{1}}{\astexpi{2}} \:\lt{a}\: \stexpprod{\astexpacci{1}}{\astexpi{2}} $}
   \DisplayProof
   & &
   \AxiomC{$\terminates{\astexpi{1}}$}
   \AxiomC{$ \astexpi{2} \:\lt{a}\: \astexpacci{2} $}
   \BinaryInfC{$ \stexpprod{\astexpi{1}}{\astexpi{2}} \:\lt{a}\: \astexpacci{2} $}
   \DisplayProof
   & &
   \AxiomC{$\astexp \:\lt{a}\: \astexpacc$}
   \UnaryInfC{$\stexpit{\astexp} \:\lt{a}\: \stexpprod{\astexpacc}{\stexpit{\astexp}}$}
   \DisplayProof
 \end{aligned}
\end{gather*}  
  \vspace*{-2ex}
  \caption{\label{fig:StExpTSS}%
    Transition system specification $\protect\StExpTSS$ for computations enabled by regular expressions.}
\end{figure}%
%
The TSS $\StExpTSS$ in Figure~\ref{fig:StExpTSS} defines,
  via derivations that it permits from its axioms, labeled transitions $\slt{\aact}$ for actions $\aact$ that occur in a regular expressions,
    and immediate successful termination via the unary predicate~$\sterminates$.
The process interpretation $\procint{\astexp}$ of a regular expression $\astexp$ is then defined
  as the sub-\LTS\ that is induced by $\astexp$ in the \LTS\ on regular expressions that is defined via derivability in~$\StExpTSS$.   
%
\begin{figure}[tb!]
\begin{center}
\begin{tikzpicture}
\matrix[anchor=north,row sep=1cm,%
        every node/.style={draw,thick,circle,minimum width=2.5pt,fill,inner sep=0pt,outer sep=2pt},%
        ampersand replacement=\&] 
        at (0,0) {
  \node(C1-0){};
  \\
  \node(C1-1){};
  \\
  \node(C1-2){};
  \\
  \node(C1-3){};
  \\      
};
\matrix[anchor=north,row sep=1cm,%
        every node/.style={draw,thick,circle,minimum width=2.5pt,fill,inner sep=0pt,outer sep=2pt},%
        ampersand replacement=\&] 
        at (4.25,-0.7) {
  \node(2E-0){};
  \\
  \node(2E-1){};
  \\
  \node(2E-2){};
  \\      
}; 
\draw[<-,thick,color=chocolate,>=latex](C1-0) -- ++ (90:0.5cm);
\draw[->] (C1-0) to node[right]{$\!{\scriptstyle a}$} (C1-1);
\draw[->] (C1-1) to node[right]{$\!{\scriptstyle a}$} (C1-2);
\draw[->] (C1-2) to node[right]{$\!{\scriptstyle b}$} (C1-3);
\draw[shorten >=3pt,shorten <=3pt] ([shift=(270:1.17cm)]C1-2) arc[radius=1.17cm,start angle=270,end angle=90];
\path (C1-2) ++ (-1.3cm,0cm) node{${\scriptstyle a}$};
\draw[->,shorten >=3pt,shorten <=3pt] ([shift=(270:1.13cm)]C1-1) arc[radius=0.565cm,start angle=270,end angle=90];
\path ([shift=(90:0.575cm)]C1-2) ++ (-0.69cm,0cm) node{${\scriptstyle b}$};
\matrix[anchor=north,row sep=1cm,%
        every node/.style={draw,thick,circle,minimum width=2.5pt,fill,inner sep=0pt,outer sep=2pt},%
        ampersand replacement=\&] 
        at (8.5,0) {
  \node(C2-0){};
  \\
  \node(C2-1){};
  \\
  \node(C2-2){};
  \\
  \node(C2-3){};
  \\      
};    
\draw[<-,thick,color=chocolate,>=latex](2E-0) -- ++ (90:0.5cm);
\draw[->] (2E-0) to node[right]{$\!{\scriptstyle a}$} (2E-1);
\draw[->] (2E-1) to node[right]{$\!{\scriptstyle a}$} (2E-2);
\draw[->,shorten >=3pt,shorten <=3pt] ([shift=(270:1.17cm)]2E-1) arc[radius=1.17cm,start angle=-90,end angle=90];
\path (2E-1) ++ (1.3cm,0cm) node{${\scriptstyle b}$};
\draw[->,out=180,in=180,distance=1.05cm] (2E-2) to node[left]{$\black{\scriptstyle b}$} (2E-1);
%
\draw[<-,thick,color=chocolate,>=latex](C2-0) -- ++ (90:0.5cm);
\draw[->] (C2-0) to node[right]{$\!{\scriptstyle a}$} (C2-1);
\draw[->] (C2-1) to node[right]{$\!{\scriptstyle a}$} (C2-2);
\draw[->] (C2-2) to node[right]{$\!{\scriptstyle b}$} (C2-3);
\draw[->,shorten >=3pt,shorten <=3pt] ([shift=(0:0pt)]C2-2) arc[radius=1.17cm,start angle=-90,end angle=90];
\path (C2-1) ++ (1.3cm,0cm) node{${\scriptstyle b}$};
\draw[->,shorten >=3pt,shorten <=3pt] ([shift=(270:1.16cm)]C2-2) arc[radius=0.58cm,start angle=-90,end angle=75];
\path ([shift=(270:0.575cm)]C2-2) ++ (0.72cm,0cm) node{${\scriptstyle a}$};
%
%
%
\draw[color=magenta,densely dashed] (C1-0) to (2E-0);
\draw[color=magenta,densely dashed] (C1-1) to (2E-1);
\draw[color=magenta,densely dashed] (C1-2) to (2E-2);
\draw[color=magenta,densely dashed] (C1-3) to (2E-0);
%
%
%
\draw[color=magenta,densely dashed] (C2-0) to (2E-0);
\draw[color=magenta,densely dashed] (C2-1) to (2E-1);
\draw[color=magenta,densely dashed] (C2-2) to (2E-2);
\draw[color=magenta,densely dashed] (C2-3) to (2E-1);

\path (C1-0) ++ (-0.7cm,0.25cm) node{\scalebox{1.25}{$\agraphi{1}$}};

\path (2E-2) ++ (-4.25cm,-1.4cm) node{$\procint{(a \cdot (a \cdot (b + b \cdot a))^*) \cdot 0}$}; 
\path (2E-2) ++ (-4.25cm,-2cm) node{$\sprocint$-expressible, hence $\procsem{\cdot}$\nb-expr.};
\path (2E-2) ++ (-4.25cm,-2.6cm) node{${(a \cdot (a \cdot (b + b \cdot a))^*) \cdot 0}$};

\path (C1-0) ++ (3.75cm,-0.15cm) node{\scalebox{1.25}{$\agraphi{0}$}};
 
\path (2E-2) ++ (-0cm,-1.4cm) node{? $\in \fap{\textit{im}}{\procint{\cdot}}$ ?};
\path (2E-2) ++ (-0cm,-2cm) node{$\procsem{\cdot}$\nb-expressible};
\path (2E-2) ++ (-0cm,-2.6cm) node{\scalebox{1.2}{$\sprocsemeq$}};

\path (C1-0) ++ (7.75cm,0.25cm) node{\scalebox{1.25}{$\agraphi{2}$}};

\path (2E-2) ++ (4.25cm,-1.4cm) node{$\procint{(1 \cdot (((a \cdot a) \cdot (b \cdot a)^*) \cdot b)^*) \cdot 0}$};
\path (2E-2) ++ (4.25cm,-2cm) node{$\sprocint$-expressible, hence $\procsem{\cdot}$\nb-expr.};
\path (2E-2) ++ (4.25cm,-2.6cm) node{$(1 \cdot (((a \cdot a) \cdot (b \cdot a)^*) \cdot b)^*) \cdot 0$};

\end{tikzpicture}
\end{center}   
  \vspace*{-2ex}
\caption{\label{fig:procsemeq}%
         Two process graphs $\protect\agraphi{1}$ and $\protect\agraphi{2}$ that are \protect\procintexpressible, and hence \protect\procsemexpressible,
         because they are the process interpretations of regular expressions as indicated. 
         $\protect\agraphi{1}$ and $\protect\agraphi{2}$ are bisimilar via bisimulations that are drawn as links $\hspace*{-0.75em}\protect\picbisimlink$
           to their joint bisimulation collapse $\agraphi{0}$
             (of which \procintexpressibility\ is at first unclear).
         It follows that also $\protect\agraphi{0}$ is \protect\procsemexpressible,
           and that process semantics equality holds between the regular expressions with interpretations $\protect\agraphi{1}$ and $\protect\agraphi{2}$, respectively.
         In this example $\agraphi{0}$ is actually also in the image of $\procint{\cdot}$, hence \protect\procintexpressible,
           as witnessed for example by $\agraphi{0} = \procint{((1 \cdot a) \cdot (a \cdot (b + b \cdot a))^*) \cdot 0}$.  
          }
\end{figure}%
%
See Figure~\ref{fig:procsemeq} for suggestive examples of (bisimilar) process interpretations of two simple regular expressions.
In process graph illustrations there and later we indicate 
  the start vertex by a brown arrow~\picarrowstart,
  and the property of a vertex $\avert$ to permit immediate successful termination
  by emphasizing $\avert$ in brown as \pictermvert\ with~a~boldface~ring. 
  
It is interesting to note that the so-defined process interpretation of regular expressions
  corresponds directly to \nondeterministic\ \finitestate\ automata (\NFAs) that are defined
    via iterations of Antimirov's partial derivatives \citeprocint{anti:1996}.%
      \footnote{Antimirov did not have a process semantics in mind, but he had set out to define, for every regular expression $\astexp$,
        an \NFA\ that is typically smaller than the deterministic automaton (\DFA) 
          as usually associated with $\astexp$ in automata and language theory.}
               
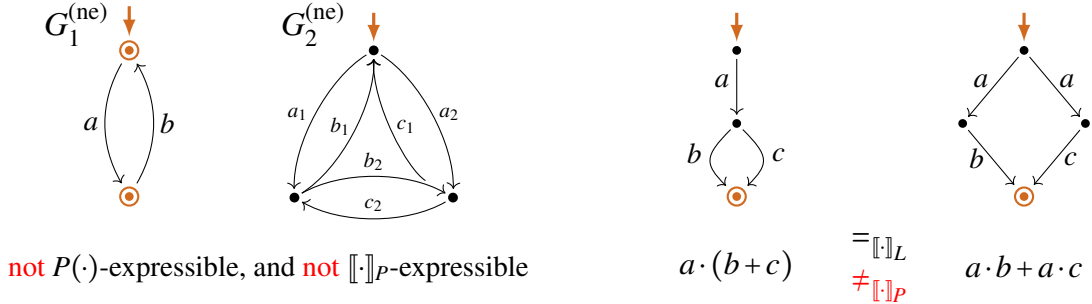
\begin{figure}[t!]
\begin{center}  
\begin{tikzpicture}


\matrix[anchor=north,row sep=0.8cm,every node/.style={draw,thick,circle,minimum width=2.5pt,fill,inner sep=0pt,outer sep=2pt}] at (0,0) {
  \node[chocolate](c-11-0){}; 
  \\
  \node[opacity=0](helper-11){};
  \\
  \node[chocolate](c-11-1){};
  \\
};

\draw[<-,very thick,>=latex,color=chocolate,shorten <= 2pt](c-11-0) -- ++ (90:0.5cm ++ 2.5pt);

\path (c-11-0) ++ (-0.7cm,0.35cm) node{\scalebox{1.15}{$\agraphbp{1}{\textnf{(ne)}}$}};

\draw[thick,chocolate] (c-11-0) circle (0.12cm);
\draw[->,bend right,shorten <= 2.5pt,shorten >= 2.5pt] (c-11-0) to node[left]{${a}\!$} (c-11-1);
                                                                
\draw[thick,chocolate] (c-11-1) circle (0.12cm); 
\draw[->,bend right,shorten <= 2.5pt,shorten >= 2.5pt] (c-11-1) to node[right]{$\!{b}$} (c-11-0);



\matrix[anchor=north,row sep=0.8cm,column sep=0.924cm,ampersand replacement=\&,
        every node/.style={draw,very thick,circle,minimum width=2.5pt,fill,inner sep=0pt,outer sep=2pt}] at (3.25,0) {
                   \& \node(c-21-0){};               \&
  \\
                   \& \node[opacity=0](helper-2-C21){}; \&                  
  \\
  \node(c-21-1){}; \& \node[opacity=0](helper-2-C21){}; \&     \node(c-21-2){};
  \\
};
\draw[<-,very thick,>=latex,color=chocolate](c-21-0) -- ++ (90:0.5cm);

\path (c-21-0) ++ (-0.8cm,0.35cm) node{\scalebox{1.15}{$\agraphbp{2}{\textnf{(ne)}}$}};

\draw[->,bend right,distance=0.6cm] (c-21-0) to node[left,xshift=0.2ex]{${\scriptstyle a_1}$} (c-21-1); 
\draw[->,bend left,distance=0.6cm,] (c-21-0) to node[right,xshift=-0.2ex]{${\scriptstyle a_2}$} (c-21-2);

\draw[->,bend right,distance=0.6cm] (c-21-1) to node[left,xshift=0.3ex,pos=0.55]{${\scriptstyle b_1}$} (c-21-0); 
\draw[->,bend left,distance=0.6cm]  (c-21-1) to node[above,yshift=-0.325ex]{${\scriptstyle b_2}$} (c-21-2); 

\draw[->,bend left,distance=0.6cm,shorten <= 9pt] (c-21-2) to node[right,xshift=-0.325ex,pos=0.55]{${\scriptstyle c_1}$} (c-21-0); 
\draw[->,bend left,distance=0.6cm]  (c-21-2) to node[above,yshift=-0.2ex]{${\scriptstyle c_2}$} (c-21-1);

%
\path (c-11-0) -- (c-21-0) node[pos=0.575](helper-3){};
\path (helper-3) ++ (0cm,-2.9cm) node{{\textcolor{red}{not} $\procint{\cdot}$\nb-expressible},
                                        and {\textcolor{red}{not} $\procsem{\cdot}$\nb-expressible}};


\matrix[anchor=north,row sep=0.8cm,column sep=0.7cm,ampersand replacement=\&,every node/.style={draw,thick,circle,minimum width=2.5pt,fill,inner sep=0pt,outer sep=2pt}] at (10,0) {
                 \&  \node(a-1){};  \&                \&[0cm]  \&[0cm]  \&                \& \node(a-2){}; 
  \\
                 \&  \node(bc-1){}; \&                \&         \&         \& \node(b-2){};  \& \node[draw=none,fill=none](bc-2){};   \& \node(c-2){};
  \\
  \node[draw=none,fill=none](t1-1){}; 
                 \&  \node[chocolate](1-1){};
                                    \& \node[draw=none,fill=none](t2-1){};   
                                                      \&         \&         \& \node[draw=none,fill=none](t1-2){}; 
                                                                                              \& \node[chocolate](1-2){};             \& \node[draw=none,fill=none](t2-2){};
  \\
};

\draw[<-,very thick,>=latex,color=chocolate](a-1) -- ++ (90:0.5cm);
\draw[->] (a-1) to node[left,pos=0.4,xshift=0.2ex]{$\aact$} (bc-1);
\draw[->,shorten >= 2.5pt,out=225,in=135,distance=0.5cm] (bc-1) to node[left,pos=0.375]{${\bact}$}  (1-1);
\draw[->,shorten >= 2.5pt,out=-45,in=45,distance=0.5cm]  (bc-1) to node[right,pos=0.4]{${\cact}$} (1-1);
%
%
\draw[thick,chocolate] (1-1) circle (0.12cm);
%

\path(bc-1) ++ (0cm,-1.9cm) node(exp-1){{\large $\stexpprod{\aact}{(\stexpsum{\bact}{\cact})}$}};

\draw[<-,very thick,>=latex,color=chocolate](a-2) -- ++ (90:0.5cm);
\draw[->] (a-2) to node[left,pos=0.4]{$\aact$} (b-2);
\draw[->] (a-2) to node[right,pos=0.4]{$\aact$} (c-2);
\draw[->,shorten >= 2.5pt] (b-2) to node[left,pos=0.475,xshift=0.15ex]{$\bact$} (1-2);
\draw[->,shorten >= 2.5pt] (c-2) to node[right,pos=0.475,xshift=-0.15ex]{$\cact$} (1-2);
%
%
\draw[thick,chocolate] (1-2) circle (0.12cm);
%

\path(bc-2) ++ (0,-1.9cm) node(exp-2){{\large $\stexpsum{\stexpprod{\aact}{\bact}}{\stexpprod{\aact}{\cact}}$}};

\path (exp-1) -- (exp-2) node[midway,yshift=1.5ex]{{\Large $\langsemeq$}};
\path (exp-1) -- (exp-2) node[midway,red,yshift=-1.5ex]{{\Large $\notprocsemeq$}};

\end{tikzpicture}
\end{center}
\vspace*{-2ex}
\caption{\label{fig:expressible:rdistr}%
         On the left: Two process graphs that are neither $\procint{\cdot}$\protect\nb-expressible
                                                (that is, not in the image of the process interpretation $\sprocint$) 
                                                nor $\procsem{\cdot}$\protect\nb-expressible
                                                (that is, not bisimilar to the process interpretation of any regular expression).
         On the right: two regular expressions with the same language semantics (associated language)
                       but different process semantics,
                       since the process interpretations are not bisimilar;
                       therefore right-distributivity does not hold for $\sprocsemeq$,
                         which entails that fewer identities hold for $\sprocsemeq$ than for $\slangsemeq$.
  }  
\end{figure}

Based on the process interpretation $\procint{\cdot}$,
  Milner then defined the process semantics of a regular expression $\astexp$
      as $\procsem{\astexp} \defdby \eqcl{\procint{\astexp}}{\sbisimsubscript}$
        where $\eqcl{\procint{\astexp}}{\sbisimsubscript}$ is the equivalence class of $\procint{\astexp}$
          with respect to bisimilarity~$\sbisim$.
In analogy to how \languagesemantics\ equality $\langsemeq$ of regular expressions 
  is defined from the language semantics $\langsem{\cdot}$
    (namely as $\astexp \langsemeq \bstexp$ if $\langint{\astexp} = \langsem{\astexp} = \langsem{\bstexp} = \langint{\astexp}$,
      for all regular expressions $\astexp$ and $\bstexp$, where $\langint{\cstexp}$ is the language defined by a regular expression $\cstexp$)
Milner was then interested in \processsemantics\ equality $\sprocsemeq$ that is defined, for all regular expressions $\astexp$ and~$\bstexp$~by:                     
  \begin{align*}
    \astexp
      \procsemeq
        \bstexp
          \;\; \funin \, 
          & \Longleftrightarrow\;\;
            \procsem{\astexp}
              =
            \procsem{\bstexp}
          \\
          & \Longleftrightarrow\;\;             
            \procint{\astexp}
              \bisim
            \procint{\bstexp} \punc{.}
  \end{align*}  
As the process interpretations of the regular expressions in Figure~\ref{fig:expressible:rdistr} are bisimilar,
  it follows that these regular expressions are linked by $\sprocsemeq$.
%
 
Milner realized in \citeprocint{miln:1984} that the process semantics $\procsem{\cdot}$ of regular expressions differs from the language semantics $\langsem{\cdot}$
  in at least two respects:
    first, $\procsem{\cdot}$ is incomplete,
      and second, \processsemantics\ equality~$\sprocsemeq$ satisfies fewer identities than \languagesemantics\ equality~$\slangsemeq$.

We start by explaining incompleteness of $\procsem{\cdot}$.
Language semantics $\langsem{\cdot}$ is complete in the following sense: 
  every language that is accepted by some finite-state automaton (FA) is the language that is defined by some regular expression;
    that is, every FA-accepted language is \emph{\langsemexpressible}.
However, an analogous statement does not hold for the process interpretation:
  not every finite process graph is `\procsemexpressible' in the sense of that it is `\procintexpressible' by a regular expression.
Here we call a finite process graph \emph{\procsemexpressible}
  if it is bisimilar to a \emph{\procintexpressible} process graph,
    by which we mean the process interpretation of some regular expression (and hence a graph in the image of $\procint{\cdot}$. 
That not every finite process graph is \procintexpressible\
  follows from the fact that there are finite process graphs that are not \procsemexpressible, either. 
Indeed, Milner proved in \citeprocint{miln:1984} that the process graph $\agraphbp{2}{\textnf{(ne)}}$ in Figure~\ref{fig:expressible:rdistr}
  not only is not \procintexpressible, but that it is not \procsemexpressible, either.  
He also conjectured that also $\agraphbp{1}{\textnf{(ne)}}$ in Figure~\ref{fig:expressible:rdistr}
  is not \procsemexpressible. That was later shown by Bosscher \citeprocint{boss:1997}.

Milner also noticed in \citeprocint{miln:1984} that some
 identities that hold for \languagesemantics\ equality $\slangsemeq$ 
   are not true any longer for process semantics equality $\sprocsemeq$.
Most notably this is the case for \rightdistributivity\ $\astexp \cdot (\bstexp + \cstexp) = \astexp \cdot \bstexp + \astexp \cdot \cstexp$,
  which is violated just as for the comparison of process terms via bisimilarity;
    see the well-known counterexample in Figure~\ref{fig:expressible:rdistr}. 
The \languagesemantics\ identity $\astexp \cdot \stexpzero = \stexpzero$ is also violated in the process semantics.
%
\begin{figure}[t!]
  \renewcommand{\prod}{\mathrel{\cdot}}
    
\begin{align*}
  &
  \begin{alignedat}{4}
    \assocstexpsum & \;\;\, & 
      e + (f + g) 
        & {} \formeq 
      (e + f) + g
      & \qquad\qquad\;\;
    \leftidstexpprod & \;\;\, & 
      e
        & {} \formeq 
      1 \prod e
      \\[-0.25ex]
    \neutralstexpsum & &
      e + 0 
        & {} \formeq 
      e
      &  
    \rightidstexpprod & &
      e
        & {} \formeq
      e \prod 1
      \\[-0.25ex]
    \commstexpsum & &
      e + f 
        & {} \formeq 
      f + e
      &
    \deadlockax & &
      0
        & {} \formeq
      0 \prod e
      \\[-0.25ex]
    \idempotstexpsum & &
      e + e 
        & {} \formeq 
      e   
      & \quad
    \recdefstexpit & & 
      e^*
        & {} \formeq
      1 + e \prod e^*
      \\[-0.25ex]
    \assocstexpprod & &
      e \prod (f \prod g) 
        & {} \formeq 
      (e \prod f) \prod g 
      &
    \termbodystexpit & \; &
        e^*
          & {} \formeq
        (1 + e)^*
      \\[-0.25ex]
    \rightdistr & &
      (e + f) \prod g 
        & {} \formeq
      e \prod g  +  f \prod g
  \end{alignedat}
  & & 
  \begin{gathered}[c]
    \renewcommand{\fCenter}{\formeq}
    \Axiom$ e \fCenter f \prod e + g $
    \RightLabel{\RSPstar\ {\small (if $\notterminates{f}$)}}
    \UnaryInf$ e \fCenter f^* \prod g $
    \DisplayProof
  \end{gathered} 
\end{align*}
\vspace*{-2.5ex}
\caption{\label{fig:milnersys}%
         Milner's equational proof system \protect\milnersys\ for process semantics equality $\protect\procsemeq$ of regular expressions
           with the fixed-point rule \protect\RSPstar\ in addition to the (not shown) basic rules for reasoning with equations
           (which guarantee that derivability in \protect\milnersys\ is a congruence relation).
         From \milnersys\ the complete proof system for language equivalence $\slangsemeq$ 
           due to Aanderaa arises
           by adding the axioms $ e \prod (f + g) = e \prod f + e \prod g $ 
             and $ e \prod 0 = 0$ (which are not sound for $\procsemeq$)
               and by dropping $\deadlockax$ (which then is derivable).}
\end{figure}
%
%
In order to define a natural sound adaptation (that we here designated by) \milnersys, see Figure~\ref{fig:milnersys}, of the complete axiom systems for $\slangsemeq$ 
  by Aanderaa \citeprocint{aand:1965} and Salomaa \citeprocint{salo:1966},
Milner dropped these two identities from Aanderaa's system, but added the sound identity $\stexpzero \cdot \astexp = \stexpzero$.   
 
These two pecularities of the process semantics led Milner to formulating two questions
  concerning recognizability of expressible process graphs,
    and axiomatizability of \processsemantics\ equality:

\begin{enumerate}[label={{\bf (E)}},leftmargin=*,align=right,labelsep=0.75ex,itemsep=0.25ex] 
      %
  \item[{\crtcrossreflabel{{\bf (E)}}[E]}]
    How can \procsemexpressible\ process graphs be characterized structurally,
      that is, those finite process graphs that are bisimilar to process interpretations of regular expressions?
  \item[{\crtcrossreflabel{{\bf (A)}}[A]}]
    Is the natural adaptation \milnersys\ to \processsemantics\ equality $\sprocsemeq$ (see Figure~\ref{fig:milnersys} for \milnersys)
      of Salomaa's and Aanderaa's complete proof systems for \languagesemantics\ equality $\slangsemeq$ 
        complete for $\sprocsemeq\,$? 
\end{enumerate}

The expressibility question \ref{E} seems to have received only limited attention at first.
The reason may have been because it asks for a structural property of (the \procsemexpressible) process graphs
  that is invariant under bisimilarity. This is a difficult aim, because bisimulations can significantly distort the topological structure
    of labeled transition graphs. 
Two variants of \ref{E} have been solved after some time:
  First, the question for a natural sufficient condition for \procsemexpressibility\ of process graphs
    was answered by Baeten and Corradini in \citeprocint{baet:corr:2005}
      by the definition of process graphs that satisfy `well-behaved' recursive specifications.
  Second, the question of whether \procsemexpressibility\ of finite process graphs is decidable
    was answered by Baeten, Corradini, and myself in \citeprocint{baet:corr:grab:2007} by giving a 
      decision procedure (unfortunately it is highly \superexponential)
      that is based on minimizing \wellbehaved\ specifications under bisimilarity.
            
For the axiomatization problem \ref{A} at first only a string of partial results have been obtained. 
In particular Milner's proof system \milnersys\ has initially been shown to be complete for $\sprocsemeq$
  for the following subclasses of regular expressions:
\begin{enumerate}[label={(\alph{*})}]
  \item{}\label{it:fokk:zant:1994}
    without $\stexpzero$ and $\stexpone$, but with binary star iteration $\stexpbit{\astexpi{1}}{\astexpi{2}}$ 
    with iteration-part $\astexpi{1}$ and exit-part $\astexpi{2}$ instead of unary star
    (Fokkink and Zantema, 1994, \citeprocint{fokk:zant:1994}),
  \item{}\label{it:fokk:1996:1997}
    with $\stexpzero$, and with iterations restricted to exit-less ones $\stexpprod{\stexpit{(\cdot)}}{\stexpzero}$
    in absence of $\stexpone$ (Fokkink, 1997, \citeprocint{fokk:1997:pl:ICALP})
      and in the presence of~$\stexpone$ (Fokkink, 1996 \citeprocint{fokk:1996:term:cycle:LGPS}),
  \item{}\label{it:corr:nico:labe:2002}
    without $\stexpzero$, and with restricted occurrences of~$\stexpone$
      (Corradini, De~Nicola, and Labella, 2002 \citeprocint{corr:nico:labe:2002}),
  \item{}\label{it:grab:fokk:2020}
    \emph{\stexponefree} expressions formed with $\stexpzero$, without $\stexpone$, but with binary iteration~$\sstexpbit$
       (G, Fokkink, 2020, \citeprocint{grab:fokk:2020:lics,grab:fokk:2020:lics:arxiv},
         also showing the completeness of a proof system by Bergstra, Bethke, and Ponse \citeprocint{berg:beth:pons:1994}).    
\end{enumerate}
While the maximal subclasses in \ref{it:corr:nico:labe:2002} and \ref{it:grab:fokk:2020} are incomparable, 
these results can be joined to apply to an encompassing class that is still a proper subclass of the regular expressions, see \citeprocint{grab:fokk:2020:lics}.
Independently of these partial results concerning completeness of Milner's system \milnersys\ for subclasses of regular expressions,
  I noticed in \citeprocint{grab:2006} that from \milnersys\ a proof system that is complete for $\procsemeq$ arises 
    when the single-equation fixed-point rule \RSPstar\ is replaced by a unique-solvability principle \USP\ for systems of guarded equations.
Also in \citeprocint{grab:2006} I formulated a coinductively motivated proof system for \processsemantics\ equality $\sprocsemeq$
  that utilizes Antimirov's partial derivatives \citeprocint{anti:1996} of regular expressions.   

The principal new idea that facilitated the partial completeness result \ref{it:grab:fokk:2020} 
  in \citeprocint{grab:fokk:2020:lics,grab:fokk:2020:lics:arxiv} of \milnersys\ for \stexponefree\ regular expressions
  consisted in formulating a natural structural condition that is sufficient (but not necessary) for \procsemexpressibility\ of process graphs:
    the \emph{Loop Existence and Elimination Condition} \LEE, and its layered form \LLEE.
This condition is based on the concept of `loop (process) graph', and an elimination process of `loop subgraphs' from a given process graph.
A process graph $\agraph$ is said to have the property \LEE\ 
  if the \nondeterministic\ iterative procedure, 
    started on $\agraph$, 
    of repeatedly eliminating loop subgraphs
    is able to obtain a process graph without an infinite behavior (that is, a graph without infinite paths and traces).
We explain the definitions in some more detail below, and provide examples.       

\smallskip
\begin{figure}[t!]
\begin{center}  
\begin{tikzpicture}

%
\matrix[anchor=center,row sep=1cm,every node/.style={draw,very thick,circle,minimum width=2.5pt,fill,inner sep=0pt,outer sep=2pt}] at (0,0) {
  \node(C1-0){}; 
  \\
  \node(C1-1){};
  \\
  \node(C1-2){};
  \\
};
\draw[<-,very thick,>=latex,color=chocolate](C1-0) -- ++ (90:0.5cm);
\draw[->](C1-0) to (C1-1);
\draw[->](C1-1) to (C1-2);


  \path (C1-2) ++ (0cm,-1cm) node{{{\colorred{\st{(L1)}}}}};

%
\matrix[anchor=center,row sep=1cm,every node/.style={draw,very thick,circle,minimum width=2.5pt,fill,inner sep=0pt,outer sep=2pt}] at (2,0) {
  \node(C2-0){}; 
  \\
  \node(C2-1){};
  \\
  \node(C2-2){};
  \\
};
\draw[<-,very thick,>=latex,color=chocolate](C2-0) -- ++ (90:0.5cm);

\draw[->,very thick,color=red] (C2-0) to node[right]{${\phantom{\aact}}$} (C2-1); 
              
\draw[->] (C2-1) to node[right]{$\hspace*{-0.65pt}{\phantom{\aact}}$} (C2-2);
             
\draw[->,color=red] (C2-1) to node[right]{$\hspace*{-0.65pt}{\phantom{\aact}}$} (C2-2);
             
\draw[->,shorten >=3pt,shorten <=3pt] ([shift=(270:1.13cm)]C2-1) arc[radius=1.13cm,start angle=-90,end angle=90];
\draw[->,out=180,in=180,distance=1.05cm] (C2-2) to node[left]{$\phantom{\bact}$} (C2-1);
  
\draw[->,color=forestgreen,shorten >=3pt,shorten <=3pt] ([shift=(270:1.13cm)]C2-1) arc[radius=1.13cm,start angle=-90,end angle=90];
\draw[->,color=red,out=180,in=180,distance=1.05cm] (C2-2) to node[left]{$\phantom{\bact}$} (C2-1);
  
\path (C2-1) ++ (1.25cm,0cm) node{${\phantom{\bact}}$};
\path ([shift=(270:0.575cm)]C2-1) ++ (-0.83cm,0cm) node{${\phantom{\bact}}$};

\path (C2-0) ++ (-0.275cm,0cm) node{${\averti{0}}$};
\path (C2-1) ++ (+0.3cm,0cm) node{${\averti{1}}$}; 
\path (C2-2) ++ (+0.3cm,-0.275cm) node{${\averti{2}}$};

\path (C2-2) ++ (1.5cm,-1cm) node{{\forestgreen{(L1)},\colorred{\st{(L2)}},\forestgreen{(L3)}}};

%
\matrix[anchor=center,row sep=0.75cm,column sep=0.924cm,ampersand replacement=\&,
        every node/.style={draw,very thick,circle,minimum width=2.5pt,fill,inner sep=0pt,outer sep=2pt}] at (4.75,0) {
  \node(C-2-1){};  \&                  \&     \node(C-2-2){};
  \\
                   \&                  \&                  
  \\
                   \& \node(C-2-3){};  \&
  \\
};
\draw[<-,very thick,>=latex,color=chocolate](C-2-1) -- ++ (90:0.5cm);  

\draw[->,bend right,distance=0.65cm] (C-2-1) to 
                                                (C-2-2);
\draw[->,bend right,distance=0.65cm,very thick,red] (C-2-1) to 
                                                                             (C-2-3); 

\draw[->,bend right,distance=0.65cm]  (C-2-2) to 
                                                 (C-2-1); 
\draw[->,bend left,distance=0.65cm,red]  (C-2-2) to 
                                                                  (C-2-3);

\draw[->,bend right,distance=0.45cm] (C-2-3) to 
                                                ($(C-2-1)+(0.25cm,-0.2cm)$);
\draw[->,bend left,distance=0.65cm,red]  (C-2-3) to 
                                                                  (C-2-2);

\matrix[anchor=center,row sep=1.75cm,column sep=1cm,every node/.style={draw,thick,circle,minimum width=2.5pt,fill,inner sep=0pt,outer sep=2pt},
        ampersand replacement=\&] at (7.25,0) {
  \node[chocolate](C3-0){}; 
  \\
  \node[chocolate](C3-1){};  
  \\
};
\draw[<-,very thick,>=latex,color=chocolate,shorten <=0.06cm](C3-0) -- ++ (90:0.56cm);

\draw[very thick,chocolate] (C3-0) circle (0.12cm);
\draw[very thick,chocolate] (C3-1) circle (0.12cm);

\draw[->,very thick,red,bend left,distance=0.6cm,shorten <=0.09cm, shorten >=0.09cm] (C3-0) to node[above]{$\phantom{\aact}$} (C3-1); 
\draw[->,forestgreen,bend left,distance=0.6cm,shorten <=0.09cm, shorten >=0.09cm]  (C3-1) to node[below]{$\phantom{\aact}$} (C3-0);

\path (C2-2) ++ (4.95cm,-1cm) node{{\forestgreen{(L1)},\forestgreen{(L2)},{\colorred{\st{(L3)}}}}};

%
\matrix[anchor=center,row sep=1cm,every node/.style={draw,very thick,circle,minimum width=2.5pt,fill,inner sep=0pt,outer sep=2pt}] at (9.75,0) { 
  \node(C5-0){}; 
  \\
  \node(C5-1){};
  \\
  \node(C5-2){};
  \\
};
\draw[<-,very thick,>=latex,color=chocolate](C5-1) -- ++ (135:0.5cm);

\draw[->,forestgreen] (C5-0) to node[right]{${\phantom{\aact}}$} (C5-1);

\draw[->,forestgreen] (C5-0) to node[right]{${\phantom{\aact}}$} (C5-1);

\draw[->,very thick,forestgreen] (C5-1) to node[right]{$\hspace*{-0.65pt}{\phantom{\aact}}$} (C5-2);

\draw[->,out=180,in=180,distance=1.05cm] (C5-2) to node[left]{$\phantom{\bact}$} (C5-1);

\draw[->,color=forestgreen,
         shorten >=3pt,shorten <=3pt] ([shift=(270:1.13cm)]C5-1) arc[radius=1.13cm,start angle=-90,end angle=90];

\draw[->,color=forestgreen,
         out=180,in=180,distance=1.05cm] (C5-2) to node[left]{$\phantom{\bact}$} (C5-1);
        
\path (C5-1) ++ (1.25cm,0cm) node{${\phantom{\bact}}$};
\path ([shift=(270:0.575cm)]C5-1) ++ (-0.83cm,0cm) node{${\phantom{\bact}}$};

\path (C5-0) ++ (-0.275cm,0cm) node{${\averti{0}}$};
\path (C5-1) ++ (+0.3cm,0cm) node{${\averti{1}}$}; 
\path (C5-2) ++ (+0.3cm,-0.275cm) node{${\averti{2}}$};

\path (C5-2) ++ (0cm,-1cm) node{{\forestgreen{(L1)},\forestgreen{(L2)},{\forestgreen{(L3)}}}};
\path (C5-2) ++ (0cm,-1.5cm) node{\forestgreen{loop graph}};

%
\matrix[anchor=center,row sep=1cm,every node/.style={draw,very thick,circle,minimum width=2.5pt,fill,inner sep=0pt,outer sep=2pt}] at (12.5,0) { 
  \node(C6-0){}; 
  \\
  \node(C6-1){};
  \\
  \node(C6-2){};
  \\
};
\draw[<-,very thick,>=latex,color=chocolate](C6-1) -- ++ (135:0.5cm);

\draw[->] (C6-0) to node[right]{${\phantom{\aact}}$} (C6-1);

\draw[->,darkcyan] (C6-0) to node[right]{${\phantom{\aact}}$} (C6-1);

\draw[->,darkcyan] (C6-1) to node[right]{$\hspace*{-0.65pt}{\phantom{\aact}}$} (C6-2);

\draw[->,out=180,in=180,distance=1.05cm] (C6-2) to node[left]{$\phantom{\bact}$} (C6-1);

\draw[->,very thick,color=darkcyan,shorten >=3pt,shorten <=3pt] ([shift=(270:1.13cm)]C6-1) arc[radius=1.13cm,start angle=-90,end angle=90];

\draw[->,out=180,in=180,distance=1.05cm] (C6-2) to node[left]{$\phantom{\bact}$} (C6-1);
        
\path (C6-1) ++ (1.25cm,0cm) node{${\phantom{\bact}}$};
\path ([shift=(270:0.575cm)]C6-1) ++ (-0.83cm,0cm) node{${\phantom{\bact}}$};

\draw[<-,very thick,>=latex,color=darkcyan](C6-2) -- ++ (270:0.5cm);

\path (C6-0) ++ (-0.275cm,0cm) node{${\averti{0}}$};
\path (C6-1) ++ (+0.3cm,0cm) node{${\averti{1}}$}; 
\path (C6-2) ++ (+0.3cm,-0.275cm) node{${\averti{2}}$};

\path (C6-2) ++ (0.15cm,-1cm) node{\darkcyan{loop subgraph}};

\path ($(C5-0)!0.5!(C6-0)$) ++ (0cm,0.25cm) node{\Large $\aloop$};
\path (C6-2) ++ (1.1cm,-0.15cm) node{\Large $\darkcyan{\aloopi{2}}$};

\end{tikzpicture}
\end{center}  
  \vspace*{-2.5ex}%
  \caption{\label{fig:exs:nonexs:loop}%
    Four process graphs (action labels ignored) 
      that violate at least one loop graph condition \protect\ref{LG:1}, \protect\ref{LG:2}, or \protect\ref{LG:3}, 
    and a loop graph $\protect\aloop$ with one of~its~loop~subgraph~$\protect\darkcyan{\protect\aloopi{2}}$.}
\end{figure}%
A process graph $\aloop$ is called a \emph{loop (process) graph} if it satisfies the following three conditions:
  \begin{enumerate}[label={(LG\arabic{*})},leftmargin=*,itemsep=0ex]
    \item{}\label{LG:1}
      There is an infinite trace from the start vertex of $\aloop$.
    \item{}\label{LG:2}  
      Every infinite trace from the start vertex $\start$ of $\aloop$ returns to $\start$. 
    \item{}\label{LG:3}
      Immediate successful termination is only possible at the start vertex of $\aloop$.
  \end{enumerate}
In such a loop graph $\aloop$, the transitions from the start vertex are called \emph{\loopentry\ transitions},
  and all other transitions are called \emph{\loopbody\ transitions}.
By a \emph{loop subgraph} of a process graph $\agraph$
  we mean a graph $\aloop$ such that with respect to a vertex $\avert$ of $\agraph$, and a \nonempty\ set $\asettrans$ of transitions of $\agraph$ that depart from $\avert$
    the following three conditions are satisfied:
  \begin{enumerate}[label={(LSG\arabic{*})},leftmargin=*,itemsep=0ex]
    \item
      $\aloop$ is a subgraph of $\agraph$ with start vertex $\avert$
        (which may be different from the start vertex $\start$ of $\agraph$).
    \item
      $\aloop$ is generated by the transitions $\asettrans$ from $\avert$
        in the following sense: $\aloop$ contains all vertices and transitions of $\agraph$ 
        that are reachable on traces  
          that start from $\avert$ via transitions in $\asettrans$, and continue onward until $\avert$ is reached again for the first time.
    \item 
      $\aloop$ is a loop graph. 
  \end{enumerate}
  In accordance with the stipulation for loop graphs,
    in such a loop subgraph $\aloop$
      the transitions in $\asettrans$ are called \emph{\loopentry} transitions of $\aloop$, and all others \emph{\loopbody} transitions of $\aloop$. 
  In Figure~\ref{fig:exs:nonexs:loop} we have gathered, on the left,
    four examples of process graphs (with action labels ignored) that are \emph{not} loop graphs:
      each of them violates one of the conditions \ref{LG:1}, \ref{LG:2}, or \ref{LG:3}.
  The paths in red indicate violations of \ref{LG:2}, and \ref{LG:3}, respectively,
    where the thicker arrows from the start vertex indicate transitions that would need to be (but are not) \loopentry\ transitions.
  However, the loop subgraph~$\aloopi{2}$ in Figure~\ref{fig:exs:nonexs:loop} is indeed a~loop~graph. 
  
  Based on these concepts, elimination of loop subgraphs is then defined as follows.
  We say that $\agraphacc$ is the result of \emph{eliminating a loop subgraph} $\aloop$ with set $\asettrans$ of \loopentry\ transitions
    \emph{from} a process graph $\agraph$,
      and denote such an elimination step by $\agraph \elimred \agraphacc$,
        if $\agraphacc$ results from $\agraph$ by first removing the transitions in $\asettrans$ 
          and by then applying garbage collection of vertices and transitions that have become unreachable from the start vertex of $\agraph$ due to the transition removals.
  See Figure~\ref{fig:ex:LEE} for an example of three loop elimination steps.
  As for \nonexamples, note that neither of two not \procsemexpressible\ graphs $\agraphbp{1}{\textnf{(ne)}}$
    and $\agraphbp{2}{\textnf{(ne)}}$ in Figure~\ref{fig:expressible:rdistr} are loop graphs, nor do they contain loop subgraphs;
  hence neither of $\agraphbp{1}{\textnf{(ne)}}$ and $\agraphbp{2}{\textnf{(ne)}}$ permits a \loopelimination\ step.
          
\input{figs/fig-ex-LEE.tex}

We say that a process graph $\agraph$ \emph{has the property \LEE} (resp.\ \emph{has the property \LLEE} (\emph{layered} \LEE))
  if there is a finite sequence of \loopelimination\ steps $\agraph \elimredrtc \agraphacc$ from $\agraph$
    such that the resulting graph $\agraphacc$ does not permit an infinite trace
      (and resp., if additionally during the elimination steps in $\agraph \elimredrtc \agraphacc$
       it never happens that a transition is removed that was a \loopbody\ transition of a loop subgraph that was eliminated in an earlier step).
It can be shown that although the property \LLEE\ is a formally stronger requirement than the property \LEE,
  which often helps to simplify proofs,
    both properties are equivalent. 
See Figure~\ref{fig:ex:LEE} for an example of a process graph $\agraph$ with the properties \LEE\ and \LLEE\
  as is witnessed there by a sequence of three loop elimination steps that lead to graph $\agraph'''$ without infinite traces. 
The not \procsemexpressible\ graphs $\agraphbp{1}{\textnf{(ne)}}$ and $\agraphbp{2}{\textnf{(ne)}}$ in Figure~\ref{fig:expressible:rdistr}
  do not satisfy \LLEE\ and \LEE, since loop elimination is not successful on them: they do not enable \loopelimination\ steps, but facilitate infinite traces.
   
The reason why the definition of the properties \LEE\ and \LLEE\ has facilitated progress concerning the problem \ref{A}
  was that they define manageable conditions that could be used for proofs
    about process graphs that are linked by functional bisimulations.
Specifically for obtaining the partial result \ref{it:grab:fokk:2020} in \citeprocint{grab:fokk:2020:lics,grab:fokk:2020:lics:arxiv} 
  it was crucial that we could prove the following facts:
  \begin{itemize}[leftmargin=*,align=left,itemsep=0ex]
    \item[{\crtcrossreflabel{{\bf (I)$_{\hspace*{0.5pt}\text{\nf \st{1}}}$}}[I-stexponefree]}]
      Process interpretations of \stexponefree\ regular expressions satisfy \LLEE\
          (see \citeprocint{grab:fokk:2020:lics:arxiv,grab:fokk:2020:lics}). 
    \item[{\crtcrossreflabel{{\bf (E)$_{\hspace*{0.5pt}\text{\nf \st{1}}}$}}[E-stexponefree]}]
      Finite process graphs with \LLEE\ are \procsemexpressible, by \stexponefree\ regular expressions
          (see \citeprocint{grab:fokk:2020:lics:arxiv,grab:fokk:2020:lics}). 
    \item[{\crtcrossreflabel{{\bf (C)}}[C]}]
      \LLEE\ is preserved along functional bisimilarity, and consequently, also by the operation of bisimulation collapse
          (see \citeprocint{grab:fokk:2020:lics:arxiv,grab:fokk:2020:lics}). 
  \end{itemize}  
  
Additionally, the property \LLEE\ permitted me to formulate a coinductive version \coindmilnersys\ of Milner's system \milnersys\
  that also permits cyclic derivations of the form of process graphs with the property \LLEE, 
    see \citeprocint{grab:2021:calco,grab:2021:calco:arxiv,grab:2023:LMCS}.
The system \coindmilnersys\ could be viewed as being located proof-theoretically half-way
  in between \milnersys\ and bisimulations between process interpretation.
As such it could be expected to form a natural beachhead for a completeness proof of \milnersys.       
  
These results raised my hope that the argumentation could be extended quite directly
  to the full set of regular expressions (including $\stexpone$ and with unary iteration instead of binary iteration)
    as well as to process graphs with \onetransitions\ and with the property \LEE.
While the generalization of \ref{I-stexponefree} to all regular expressions does not hold,
  this obstacle could be overcome by defining a refined process interpretation with the desired property.
Together with a rather straightforward generalization of \ref{E-stexponefree} we obtained:        
  \begin{itemize}[leftmargin=*,align=left,itemsep=0ex]
    \item[{\crtcrossreflabel{{\st{\bf (I)}}}[not-I]}]
      The process interpretation $\procint{\astexp}$ of a regular expression $\astexp$ 
        does not always satisfy \LLEE\ (nor \LEE) (see \citeprocint{grab:2020:termgraph-report,grab:2021:termgraph-postproceedings}).
    \item[{\crtcrossreflabel{{\bf (RI)$_{\sone}$}}[RI-one]}]    
      There is a refined process interpretation $\procintone{\cdot}$
        that produces finite process graphs with \onetransitions\
          such that, for every regular expression $\astexp$,
            $\procintone{\astexp}$ satisfies \LLEE,
            $\procintone{\astexp}$ is a refinement of $\procint{\astexp}$ by sharing transitions by means of added \onetransitions,
            and $\procintone{\astexp} \onebisim \procint{\astexp}$, that is, $\procintone{\astexp}$ is bisimilar to $\procint{\astexp}$
              when \onetransitions\ are interpreted as empty steps
                (see \citeprocint{grab:2023:i-pi-not-closed-bc:arxiv},
                 and a slightly weaker statement in \citeprocint{grab:2020:termgraph-report,grab:2021:termgraph-postproceedings}). 
              
    \item[{\crtcrossreflabel{{\bf (E)$_{\sone}$}}[E-one]}]   
      Finite process graphs with \onetransitions\ and with \LLEE\ are \procsemexpressible.
        (See \citeprocint{grab:2021:calco:arxiv,grab:2021:calco,grab:2023:LMCS}.)
  \end{itemize}
However, critically, a direct generalization of our argument broke down dramatically due to the fact that the collapse statement \ref{C}
  did not generalize to process graphs with \LLEE\ that contain \onetransitions:
  \begin{itemize}[leftmargin=*,align=left,itemsep=0ex]
    \item[{\crtcrossreflabel{{\st{{\bf (C)}$_{\sone}$}}}[not-C-one]}]
      \LLEE\ is \underline{\smash{not preserved}} under bisimulation collapse of process graphs with \onetransitions.
        A counterexample holds for the process graph $\agraph$ on the left in Figure~\ref{fig:twincrystal}.
          (See \citeprocint{grab:2022:lics,grab:2022:lics:arxiv,grab:2023:i-pi-not-closed-bc:arxiv}.)
  \end{itemize}  
As a consequence of this statement the image of the process interpretation is \underline{not closed} under bisimulation,
  see \citeprocint{grab:2023:i-pi-not-closed-bc:arxiv}.
This, however, contrasts with the image of a `compact' version of the process interpretation that,
  when restricted to `\txtusof' regular expressions, \underline{is closed} under bisimulation collapse, 
    see \citeprocint{grab:2023:i-pi-not-closed-bc:arxiv,grab:2024:TERMGRAPH}.
Now due to \ref{not-C-one} the proof strategy we used in \citeprocint{grab:fokk:2020:lics,grab:fokk:2020:lics:arxiv}
  for showing completeness of \milnersys\ for \stexponefree\ regular expressions,
    turned out not to work for showing completeness of \milnersys\ for the full class of regular expressions.
At the very least it was in need of a substantial refinement.    
    
\input{figs/fig-twincrystal.tex}%
What came to my rescue here was that the counterexample for \LLEEpreserving\ collapse of process graphs with \onetransitions\ and \LLEE, 
  the graph~$\agraph$ in \ref{fig:twincrystal}, is of a specific symmetric form. 
It is a \emph{\twincrystal}, a process graph with \onetransitions\ and with \LLEE\
  that is \nearcollapsed\ in the sense that non-identical bisimilar vertices appear only as pairs.
More precisely, \twincrystals\ are process graphs with \onetransitions\ and with \LLEE\ 
  that consist of a single strongly connected component (\scc), 
    and of two parts, the top-part and the pivot-part (see in Figure~\ref{fig:twincrystal} on the right).
Each part by itself is bisimulation collapsed, 
  and hence any two bisimilar vertices in the \twincrystal\ must occur in different of the top and pivot parts,
    and are linked by a \selfinverse\ (partial) counterpart function.
Process graphs with \onetransitions\ and with \LLEE\ that are collapsed apart from within \sccs,
  and in which all \sccs\ are either collapsed or \twincrystals, we called \emph{crystallized.}
For this concept it was possible to show:
 \begin{itemize}[leftmargin=*,align=left,itemsep=0ex]
    \item[{\crtcrossreflabel{{{\bf (NC)}$_{\sone}$}}[NC]}]
      Every finite process graph with \onetransitions\ and with \LLEE\
        can be minimized under bisimilarity to obtain a crystallized process graph 
          (see \citeprocint{grab:2022:lics,grab:2022:lics:arxiv,grab:2022:lics:poster}).
  \end{itemize}    
This statement is based on an effective \emph{crystallization procedure} of process graphs with \LLEE\ and with \onetransitions:
  it minimizes all \sccs\ of the graph either to \twincrystals\ or collapsed parts of the graph,
    and also guarantees that the resulting graph is collapsed apart from within those \sccs\ that are \twincrystals.
The symmetric structure of \twincrystals\ can then be used to show that self-bisimulations of crystallized process graphs 
  are of a particularly easy kind, which can be assembled from bisimulation slices that act on the \twincrystal-\sccs\
    \citeprocint{grab:2021:NWPT}.    
This result on crystallized versions of process interpretations permitted me to adapt the proof strategy that Fokkink and I had used previously
  to also show completeness of \milnersys\ for $\sprocsemeq$ on the full class of regular expressions, 
    see \citeprocint{grab:2022:lics,grab:2022:lics:arxiv}, and the poster~\citeprocint{grab:2022:lics:poster}.   
    
There is now much hope that the crystallization technique that we developed for solving the axiomatization question \ref{A}
  may turn out to facilitate also significant improvements for answers to the expressibility question \ref{E}.
We return to the expressibility question \ref{E} at the end of the next section.

\bibliographystyleprocint{eptcs}
\bibliographyprocint{procint}

\section{Current and Future Work}%
  \label{future}

This section touches on my current research, and lists as well as briefly motivates three research questions and projects
  that have developed out of the work that we summarized in the previous two sections.
This is organized in two subsections below that refer to the topics of Section~\ref{lambdaletrec} and Section~\ref{procint}, respectively.  

\subsection{Maximal Sharing at Run Time}

Apart from using the maximal-sharing method for functional programs 
  as a \staticanalysis\ based optimization transformation during compilation,
    one of the ideas
      for applications that Rochel and I gathered in \citelambdaletrec{grab:roch:2014:ICFP}
        was that maximal sharing could be used as an optimization transformation also repeatedly at run-time.
Making that idea fruitful, however, requires
  that representations of programs that are used in graph evaluators 
    can be linked closely with \lambdatermgraph\ representations of \lambdaletrecterms\ 
      on which the maximal-sharing method operates. 
This is necessary because graph evaluators in implementations of functional languages
  typically use supercombinator representations of \lambdaletreccalterms,
    and much computational overhead is to be expected in transformations to and from \lambdatermgraphs.
      Yet any such overhead is highly undesirable during program execution. 
Now supercombinator reduction as carried out by graph evaluators
  intuitively corresponds to \emph{\scopesharing} forms of \betareduction.%
    \footnote{Note that \scopesharing\ is distinct from the \emph{\contextsharing} forms of graph reduction 
                on which implementations of \emph{parallel} or \emph{optimal \betareduction} are based.}
And so, since \lambdatermgraphs\ contain neatly described scopes of \lambdaabstractions,
  the implementation of a \scopesharing\ form of evaluation on \lambdatermgraphs\ is conceivable.
These considerations lead me to the following research question.

\begin{resque}\label{resque:coupling:maxsharing:2:evaluation}
  Coupling of maximal sharing with evaluation, generally, and more specifically:
  \begin{enumerate}[label={(\roman{*})},itemsep=0ex]
    \item{}\label{it:1:resque:coupling:maxsharing:2:evaluation}  
      Can the maximal-sharing method for terms in the \lambdacalculus\ with \txtletrec\ 
        be coupled naturally with an efficient evaluation method (such as a standard graph-evaluation implementation)?
    \item{}\label{it:2:resque:coupling:maxsharing:2:evaluation}  
      Do \lambdatermgraphs\ (which represent \lambdaletreccalterms)
        permit a representation as interaction nets or as port graphs \citefuture{stew:2002}
          for which a form of \betareduction\ can be defined
            that preserves both \lambdatermgraph\ form and represented \lambdaabstraction\ scopes by adequately chosen multi-steps of interactions?
  \end{enumerate}
\end{resque}

In communication after the workshop, 
  Ian Mackie pointed me to his 
    interaction-net based implementation \citefuture{mack:1998,mack:2004} of an evaluation method for the \lambdacalculus .
I am grateful for this reference, first, because
  this interaction-net representation of \lambdaterms\ 
    bears a close resemblance with \lambdatermgraphs,
      and second, because it provides a mechanism for implementing scope-preserving forms of \betareduction.
Nevertheless it remains a challenging question to relate the two formalisms
  (\lambdatermgraphs\ and interaction-net representations of \lambdaterms\ in \citefuture{mack:1998})
    closely together.
Yet an interaction-net representation of \lambdatermgraphs\ close to the representation of \lambdaterms\ as used in \citefuture{mack:1998}
  seems to me to be a plausible and promising in-road
   for approaching part~\ref{it:2:resque:coupling:maxsharing:2:evaluation} of Research Question~\ref{resque:coupling:maxsharing:2:evaluation}.    

Regarding graph evaluators that implement \scopepreserving\ forms of \betareduction\
  it will also be important to explore 
    correspondences on the rewrite-step level 
      between graph evaluation steps 
        and steps of, on the one hand, \leftmostoutermost\ \betareduction, and on the other hand, reduction on super-combinator representations of \lambdaterms.
Such connections were recently outlined by van~Oostrom in \citefuture{oost:2024}:
  a correspondence between \betareduction\ steps and \combinatorreduction\ steps,
    as well as a correspondence between, \betareduction\ and \combinatorreduction\ \multisteps\ on the one hand, and \graphrewriting\ \multisteps\ on the other hand.
In doing so, van~Oostrom carried a decisive step further
  the idea that I suggested in \citefuture{grab:2019:lindepthincrease:arxiv} of using supercombinator representations for obtaining an alternative proof, based on graph-rewriting on supercombinator representations of \lambdaterms,
    of an invariance result for \leftmostoutermost\ \betareduction\ by Accattoli and Dal~Lago in \citefuture{acca:lago:2016}.
    Namely, of the result that \leftmostoutermost\ \betareduction\ in the \lambdacalculus\
      can be implemented on every reasonable machine with only a polynomial overhead in the number of computation steps. 
In preparation for such a proof 
  I had used supercombinator representations of \lambdaterms\ in \citefuture{grab:2019:lindepthincrease:arxiv} 
    for developing the result that the depth increase along any \leftmostoutermost\ \betareduction\ sequence from a \lambdaterm\
      is always bounded linearly in the number of steps of the sequence.
      
%
%
%

\subsection{Crystallization: Proof Verification, and Application to the Expressibility Problem}

Currently I am writing two articles that will provide the details of the completeness proof of Milner's proof system \milnersys.
The first article will explain the motivation of the crystallization process for process interpretations of regular expressions:
  a limit to minimization under bisimilarity of \procintexpressible\ process graphs. 
    This limit will be established specifically for the process graph $\agraph$ in Figure~\ref{fig:twincrystal} with \onetransitions.
The second article will detail the crystallization procedure 
  by which process graphs with the property \LLEE\ (which are \procsemexpressible) are minimized under bisimulation
     to obtain process graphs with \LLEE\ that are close to their bisimulation collapse.
This central result will then be used, as explained in \citeprocint{grab:2022:lics},
  to show that Milner's proof system \milnersys\ is complete with respect to process semantics equality $\procsemeq$.     
    
This completeness proof can be explained with clear conceptual concepts, and with convincing details,
  answering Milner's question \ref{A} positively.
However,  
    a verification of the crystallization procedure and the completeness proof of \milnersys\ with respect to $\procsemeq$ forms an important goal for me.    

\begin{respro} 
  Formalization of the proofs for crystallization, and completeness of \milnersys:
  \begin{enumerate}[label={(\alph{*})}]
    \item
      Develop formalizations of structure constraints for process graphs
        in order to verify the correctness of the crystallization procedure for process graphs with \LEE\ by a proof assistant.
    \item
      Use the correctness proof of crystallization to verify the completeness proof of Milner's proof system \milnersys\
        by a proof assistant.    
  \end{enumerate}  
\end{respro}

Separately I am working out a proof of the fact that the loop existence and elimination property \LEE\
  (and equivalently \LLEE) can be decided in polynomial time.
  For this result the observation is crucial that loop elimination $\selimred$ can be completed to obtain a confluent rewrite system (which is obviously terminating).
As a consequence of the efficient decidability of \LLEE\ it follows that the restriction of the expressibility problem \ref{E}
  to expressibility by regular expressions that are \txtusof\ 
    (but with unary iteration, see \citeprocint{grab:2023:i-pi-not-closed-bc:arxiv,grab:2024:TERMGRAPH}) 
      can be solved efficiently.
This is because the methods and results in \citeprocint{grab:fokk:2020:lics,grab:fokk:2020:lics:arxiv} 
  permit to show that
    a finite process graph $\agraph$ is \procsemexpressible\ by a regular expression that is \txtusof\
      if and only if 
    the bisimulation collapse of $\agraph$ satisfies \LLEE. 
Then it follows that expressibility of finite process graphs by regular expressions that are \txtusof\ can be decided in polynomial time.

The crystallization procedure that we use in the completeness proof of \milnersys\ with respect to $\sprocsemeq$
  suggests that an extension of this characterization statement to one for \procsemexpressibility\ in full generality
    is conceivable. We formulate that as our final research question.

\begin{resque}
  Is the problem of whether a finite process graph is \procsemexpressible\ efficiently decidable? 
  That is, is there a polynomial decision algorithm for it?
    Or is \procsemexpressibility\ at least fixed-parameter tractable (in \FPT) for interesting parameterizations?
\end{resque}

\bibliographystylefuture{eptcs}
\bibliographyfuture{future}


\end{document}